\documentclass[12pt]{article}
\pdfoutput=1 % if your are submitting a pdflatex (i.e. if you have
\usepackage[utf8]{inputenc}

\usepackage{amsmath,amsfonts,amssymb,epsfig}
\usepackage{slashed}
\usepackage{cite}
\usepackage{multirow}
\usepackage{multicol}
\usepackage{cite}

\usepackage{xcolor}
\usepackage[
   bookmarks=true,
   linktocpage=true,
   colorlinks=true,
   allbordercolors=white, allcolors=blue
]{hyperref}

\usepackage{calc}
\usepackage{rotating}
\usepackage[english]{babel}
\usepackage{pifont}

\usepackage{graphicx}
\usepackage{subfig}
\usepackage{float}
\usepackage{amsmath}
\usepackage{amssymb}
\usepackage{amsthm}
\usepackage{amsfonts}
\usepackage{latexsym}
\usepackage{dcolumn}
\usepackage{geometry}
\usepackage{mciteplus}
\usepackage{fancyhdr}

\usepackage[normalem]{ulem}
\usepackage[utf8]{inputenc}
\usepackage{caption}
\usepackage{array}

\def\gtwid{\mathrel{\raise.3ex\hbox{$>$\kern-.75em\lower1ex\hbox{$\sim
$}}}}
\def\vio{\mathrel{\hbox{$E$\kern-.60em\hbox{$/
$}}}}
\newcolumntype{P}[1]{>{\centering\arraybackslash}p{#1}}

\newcommand{\email}[1]{\href{mailto:#1}{#1}}
\newcommand{\newc}{\newcommand*}

\long\def\begincomment#1\endcomment{%
        \begingroup\sf\baselineskip12pt#1\endgroup}

\newc{\etal}{\textrm{et al.}} 
\newc{\eg}{\textrm{e.g.}} 
\newc{\ie}{\textrm{i.e.}}
\newc{\etc}{\textrm{etc}}
\newc\vs{\textrm{vs.}}
\newc{\cl}{\rm {C.L.}}
\newc{\ev}{\ensuremath{\,\mathrm{eV}}}
\newc{\kev}{\ensuremath{\,\mathrm{keV}}}
\newc{\mev}{\ensuremath{\,\mathrm{MeV}}}
\newc{\gev}{\ensuremath{\,\mathrm{GeV}}}
\newc{\tev}{\ensuremath{\,\mathrm{TeV}}}
\newc{\MeV}{\mev} 
\newc{\TeV}{\tev}
\newc{\invpb}{\ensuremath{/\text{pb}}}
\newc{\invfb}{\ensuremath{\,\textrm{fb}^{-1}}}
\newc\nb{\ensuremath{\,\mathrm{nb}}} \newc\pb{\ensuremath{\,\mathrm{pb}}} \newc\fb{\ensuremath{\,\mathrm{fb}}}
\newc\pc{\ensuremath{\,\mathrm{pc}}}
\newc\kpc{\ensuremath{\,\mathrm{kpc}}}
\newc\mpc{\ensuremath{\,\mathrm{Mpc}}}
\newc\ps{\ensuremath{\,\mathrm{ps}}} 
\newc\cmeter{\ensuremath{\,\mathrm{cm}}} 
\newc\meter{\ensuremath{\,\mathrm{m}}} 
\newc\kmeter{\ensuremath{\,\mathrm{km}}}
\newc\second{\ensuremath{\,\mathrm{s}}}
\newc\msecond{\ensuremath{\,\mathrm{ms}}}
\newc\nsecond{\ensuremath{\,\mathrm{ns}}}
\newc\psecond{\ensuremath{\,\mathrm{ps}}}
\newc{\chisqmin}{\ensuremath{\chi^2_{\mathrm{min}}}}
\newc{\Delchisq}{\ensuremath{\Delta\chi^2}}
\newc{\chisq}{\ensuremath{\chi^2}}
\newc{\like}{\ensuremath{\mathcal{L}}}
\newc\lsim{\ensuremath{\mathrel{\rlap{\lower4pt\hbox{\hskip1pt$\sim$}}\raise1pt\hbox{$<$}}}}
\newc\gsim{\ensuremath{\mathrel{\rlap{\lower4pt\hbox{\hskip1pt$\sim$}}\raise1pt\hbox{$>$}}}}
\newc{\VEV}[1]{\ensuremath{\langle #1 \rangle}}
\newc{\dl}{\ensuremath{\stackrel{\leftarrow}{D}}}
\newc{\dr}{\ensuremath{\stackrel{\rightarrow}{D}}}
\newc{\bcenter}{\begin{center}}    \newc{\ecenter}{\end{center}}
\newc{\bfl}{\begin{flushleft}}    \newc{\efl}{\end{flushleft}}
\newc{\bfr}{\begin{flushright}}    \newc{\efr}{\end{flushright}}

\newc{\bi}{\begin{itemize}}
\newc{\ei}{\end{itemize}}
\newc{\bed}{\begin{description}}
\newc{\eed}{\end{description}}
\newc{\ben}{\begin{enumerate}}
\newc{\een}{\end{enumerate}}

\newc{\be}{\begin{equation}}
\newc{\ee}{\end{equation}}
\newc{\bea}{\begin{eqnarray}}
\newc{\eea}{\end{eqnarray}}
\newc{\bfle}{\begin{flalign}}
\newc{\efle}{\end{flalign}}
\newc{\ra}{\rightarrow}
\newc{\alphas}{\ensuremath{\alpha_s}}
\newc{\alphatwo}{\ensuremath{\alpha_2}}
\newc{\alphaone}{\ensuremath{\alpha_1}}
\newc{\alphai}[1]{\ensuremath{\alpha_{#1}}}
\newc{\alphaem}{\ensuremath{\alpha_{\mathrm{em}}}}
\newc{\alphaeff}{\ensuremath{\alpha_{\mathrm{eff}}}}
\newc{\sineff}{\ensuremath{\sin \theta_{\mathrm{eff}}}}
\newc{\sinsqeff}{\ensuremath{\sin^2 \theta_{\mathrm{eff}}}}
\newc{\dalphahad}{\ensuremath{\Delta \alpha_{\mathrm{had}}}}
\newc{\yt}{\ensuremath{h_t}} \newc{\yb}{\ensuremath{h_b}} \newc{\ytau}{\ensuremath{h_{\tau}}}
\newc\mz{\ensuremath{M_Z}} 
\newc\mw{\ensuremath{m_W}}
\newc\mZ{\mz}        \newc\mW{\mw}
\newc\mhsm{\ensuremath{ m_{H_{\mathrm{SM}}}}}
\newc{\mtop}{\ensuremath{ m_t}}               \newc{\mtpole}{\ensuremath{ M_t}}
\newc{\mbottom}{\ensuremath{ m_b}} 
\newc{\mtau}{\ensuremath{ m_{\tau}}}
\newc{\mt}{\mtpole}
\newc{\mb}{\mbottom} 
\newc{\rtwogg}{\ensuremath{R_{h_2}(\gamma\gamma)}}
\newc{\rtwozz}{\ensuremath{R_{h_2}(ZZ)}}
\newc{\ronegg}{\ensuremath{R_{h_1}(\gamma\gamma)}}
\newc{\ronezz}{\ensuremath{R_{h_1}(ZZ)}}
\newc{\rsiggg}{\ensuremath{R_{h_\textrm{sig}}(\gamma\gamma)}}
\newc{\rsigzz}{\ensuremath{R_{h_\textrm{sig}}(ZZ)}}
\newc{\llbar}{\ensuremath{\ell\bar{\ell}}}
\newc{\tauptaum}{\ensuremath{ \tau^+\tau^-}}
\newc{\qqbar}{\ensuremath{ q\bar{q}}} \newc{\ppbar}{\ensuremath{ p\bar{p}}}
\newc{\bbbar}{\ensuremath{ b\bar{b}}} \newc{\ttbar}{\ensuremath{ t\bar{t}}}
\newc{\ffbar}{\ensuremath{ f\bar{f}}} \newc{\tautaubar}{\ensuremath{ \tau\bar{\tau}}}

\newc{\mchi}{\ensuremath{m_\neutone}}
\newc{\squark}{\ensuremath{\tilde{q}}}
\newc{\slepton}{\ensuremath{\tilde{l}}}
\newc{\gluino}{\ensuremath{\tilde{g}}} 
\newc{\mgluino}{\ensuremath{{m_{\gluino}}}}
\newc{\wino}{\ensuremath{\tilde{W}}} 
\newc{\mwino}{\ensuremath{{m_{\wino}}}}
\newc{\tone}{\ensuremath{{\tilde{t}_1}}}
\newc{\bone}{\ensuremath{{\tilde{b}_1}}}
\newc{\Hone}{\ensuremath{{\tilde{H}_{1}}}}
\newc{\Htwo}{\ensuremath{{\tilde{H}_{2}}}}
\newc{\Hhtwo}{\ensuremath{{H_{2}}}}
\newc{\qli}{\ensuremath{{\tilde{Q}_{i}}}}
\newc{\uri}{\ensuremath{{\tilde{u}_{i}}}}
\newc{\dri}{\ensuremath{{\tilde{d}_{i}}}}
\newc{\lli}{\ensuremath{{\tilde{L}_{i}}}}
\newc{\eri}{\ensuremath{{\tilde{e}_{i}}}}
\newc{\sthw}{\ensuremath{ \sin\theta_W}}              \newc{\cthw}{\ensuremath{\cos\theta_W}}
\newc{\tanthw}{\ensuremath{ \tan\theta_W}}              \newc{\cotthw}{\ensuremath{\cot\theta_W}}
\newc{\ssqthw}{\ensuremath{\sin^2 \theta_W}}
\newc{\msbar}{\ensuremath{\overline{MS}}} \newc{\drbar}{\ensuremath{\overline{DR}}}
\newc{\mtmtsmmsbar}{\ensuremath{ m_t(m_t)^{\msbar}_{{\mathrm{SM}}}}}
\newc{\mtmtsmdrbar}{\ensuremath{ m_t(m_t)^{\drbar}_{{\mathrm{SM}}}}}
\newc{\mtmtmssmdrbar}{\ensuremath{ m_t(m_t)^{\drbar}_{{\mathrm{SUSY}}}}}
\newc{\mbmbmsbar}{\ensuremath{ m_b(m_b)^{\msbar} }}
\newc{\mbmbsmmsbar}{\ensuremath{ m_b(m_b)^{\msbar}_{{\mathrm{SM}}}}}
\newc{\mbmzsmmsbar}{\ensuremath{ m_b(\mz)^{\msbar}_{{\mathrm{SM}}}}}
\newc{\mbmzsmdrbar}{\ensuremath{ m_b(\mz)^{\drbar}_{{\mathrm{SM}}}}}
\newc{\mbmzmssmdrbar}{\ensuremath{ m_b(\mz)^{\drbar}_{{\mathrm{SUSY}}}}}
\newc{\mtaumzsmmsbar}{\ensuremath{ m_{\tau}(\mz)^{\msbar}_{{\mathrm{SM}}}}}
\newc{\mtaumzsmdrbar}{\ensuremath{ m_{\tau}(\mz)^{\drbar}_{{\mathrm{SM}}}}}
\newc{\mtaumzmssmdrbar}{\ensuremath{ m_{\tau}(\mz)^{\drbar}_{{\mathrm{SUSY}}}}}
\newc{\alphasmzms}{\ensuremath{\alpha_s(M_Z)^{\overline{MS}}}}
\newc{\alphaimzms}[1]{\ensuremath{\alpha_{#1}(M_Z)^{\overline{MS}}}}
\newc{\alphaemmz}{\ensuremath{\alpha_{\mathrm{em}}(M_Z)^{\overline{MS}}}}
\newc{\mzero}{\ensuremath{{m_0}}}
\newc{\mhalf}{\ensuremath{ m_{1/2}}}
\newc{\tanb}{\ensuremath{\tan\beta}}
\newc{\azero}{\ensuremath{ A_0}}
\newc{\signmu}{\ensuremath{\rm{sgn}\,\mu}}
\newc{\atau}{\ensuremath{{A_{\tau}}}}
\newc{\mueff}{\ensuremath{\mu_{\rm{eff}}}}
\newc{\lam}{\ensuremath{{\lambda}}}
\newc{\kap}{\ensuremath{{\kappa}}}
\newc{\alam}{\ensuremath{{A_{\lambda}}}}
\newc{\akap}{\ensuremath{{A_{\kappa}}}}
\newc{\hs}{\ensuremath{ H_s}}      
\newc{\mhs}{\ensuremath{ m_{H_s}}} 
\newc{\mgut}{\ensuremath{ M_{\rm GUT}}}
\newc{\gut}{\ensuremath{{\rm GUT}}}
\newc{\mplanck}{\ensuremath{ M_{\rm P}}}      \newc{\mpl}{\ensuremath{ M_{\rm Pl}}}
\newc{\msusy}{\ensuremath{ M_{\rm SUSY}}}      \newc{\ms}{\ensuremath{ M_{\rm S}}}
 \newc{\hu}{\ensuremath{ H_u}}       \newc{\hd}{\ensuremath{ H_d}}
 \newc{\mhu}{\ensuremath{ m_{H_u}}}       \newc{\mhd}{\ensuremath{ m_{H_d}}}
 \newc{\mhuew}{\ensuremath{ m^{\ast}_{H_u}}}       \newc{\mhdew}{\ensuremath{ m^{\ast}_{H_d}}}
 \newc{\mhuewsq}{\ensuremath{ m^{\ast\, 2}_{H_u}}}       \newc{\mhdewsq}{\ensuremath{ m^{\ast\, 2}_{H_d}}}
 \newc{\mhl}{\ensuremath{m_\hl}} 
 \newc{\mhone}{\ensuremath{m_{h_1}}} 
 \newc{\mhtwo}{\ensuremath{m_{h_2}}} 
 \newc{\mhi}{\ensuremath{m_{\tilde{h}}}} 
 \newc{\mul}{\ensuremath{m_{\tilde{u}_L}}} 
 \newc{\mbone}{\ensuremath{m_{\tilde{b}_1}}}  
 \newc{\mtone}{\ensuremath{m_{\tilde{t}_1}}} 
 \newc{\ma}{\ensuremath{m_A}} 
 \newc{\mH}{\ensuremath{m_H}} 
 \newc{\maone}{\ensuremath{m_{a_1}}} 
 \newc{\matwo}{\ensuremath{m_{a_2}}}
 \newc{\hone}{\ensuremath{h_1}}
 \newc{\htwo}{\ensuremath{h_2}}
 \newc{\aone}{\ensuremath{a_1}}
 \newc{\atwo}{\ensuremath{a_2}}
 \newc{\mqthree}{\ensuremath{m_{\tilde{Q}_3}^2}}
 \newc{\muthree}{\ensuremath{m_{\tilde{u}_3}^2}}
 \newc{\mqli}{\ensuremath{m_{\tilde{Q}_{i}}}}
 \newc{\muri}{\ensuremath{m_{\tilde{u}_{i}}}}
 \newc{\mdri}{\ensuremath{m_{\tilde{d}_{i}}}}
 \newc{\mlli}{\ensuremath{m_{\tilde{L}_{i}}}}
 \newc{\meri}{\ensuremath{m_{\tilde{e}_{i}}}}
 \newc{\ts}{\ensuremath{T_{SUSY}}}

\newc{\sigsip}{\ensuremath{\sigma^{\rm SI}_{p}}}	\newc{\sigsin}{\ensuremath{\sigma^{\rm SI}_{n}}}
\newc{\sigsdp}{\ensuremath{\sigma^{\rm SD}_{p}}}	\newc{\sigsdn}{\ensuremath{\sigma^{\rm SD}_{n}}}
\newc{\sigsi}{\ensuremath{\sigma^{\rm SI}}}	\newc{\sigsd}{\ensuremath{\sigma^{\rm SD}}}
\newc{\abund}{\ensuremath{ \Omega h^2}}
\newc{\omegadm}{\ensuremath{ \Omega_{{\rm DM}}}}     \newc{\abunddm}{\ensuremath{ \Omega_{{\rm DM}} h^2}} 
\newc{\omegam}{\ensuremath{ \Omega_{{\rm m}}}}       \newc{\abundm}{\ensuremath{ \Omega_{{\rm m}} h^2}}
\newc{\omegab}{\ensuremath{ \Omega_{{\rm b}}}}	\newc{\abundb}{\ensuremath{ \Omega_{{\rm b}} h^2}}
\newc{\omegatot}{\ensuremath{ \Omega_{{\rm TOT}}}}
\newc{\omegacdm}{\ensuremath{ \Omega_{{\rm CDM}}}}   \newc{\abundcdm}{\ensuremath{ \Omega_{{\rm CDM}} h^2}}
\newc{\omegalambda}{\ensuremath{ \Omega_{\Lambda}}} \newc{\abundlambda}{\ensuremath{ \Omega_{\Lambda} h^2}}
\newc{\omegarad}{\ensuremath{ \Omega_{{\rm rad}}}}  \newc{\abundrad}{\ensuremath{ \Omega_{{\rm rad}} h^2}}
\newc{\rhocrit}{\ensuremath{ \rho_{\rm crit}}}
\newc{\rhochi}{\ensuremath{ \rho_{\chi}}}
\newc{\abunchi}{\ensuremath{\Omega_\chi h^2}}
\newc{\abundlsp}{\ensuremath{\Omega_{\rm LSP}h^2}}
\newc{\amu}{\ensuremath{ a_{\mu}}}        \newc{\amususy}{\ensuremath{ a_{\mu}^{\mathrm{SUSY}}}}
\newc{\amuexpt}{\ensuremath{ a_{\mu}^{\mathrm{expt}}}}        \newc{\amusm}{\ensuremath{ a_{\mu}^{\mathrm{SM}}}}
\newc\deltaamu{\ensuremath{\Delta a_{\mu}}} \newc{\deltaamususy}{\ensuremath{\delta a_{\mu}^{\mathrm{SUSY}}}}
\newc\gmtwo{\ensuremath{ (g-2)_{\mu}}} 
\newc{\deltagmtwomususy}{\ensuremath{\delta\left(g-2\right)_{\mu}^{\mathrm{SUSY}}}}
\newc{\deltagmtwomu}{\ensuremath{\delta\left(g-2\right)_{\mu}}}
\newc\BR{\ensuremath{\rm BR}}
\newc\bsgamma{\ensuremath{ b\rightarrow s \gamma }}
\newc\bxsgamma{\ensuremath{\overline{B}\rightarrow X_{s}\gamma}}
\newc\brbsgamma{\ensuremath{\BR\left(\bsgamma\right)}}
\newc\brbxsgamma{\ensuremath{\BR\left(\bxsgamma\right)}}
\newc\bsmumu{\ensuremath{B_s\to\mu^+\mu^-}}
\newc\brbsmumu{\ensuremath{\BR\left(B_s\to\mu^+\mu^-\right)}}
\newc\bdmmumu{\ensuremath{\overline{B}_d\to\mu^+\mu^-}}
\newc\bbbarmix{\ensuremath{\overline{B}_s\mbox{-}B_s}}      % B_s mixing
\newc\delmbs{\ensuremath{\Delta M_{B_s}}}
\newc{\butaunu}{\ensuremath{B_u \rightarrow \tau \nu}}
\newc{\brbutaunu}{\ensuremath{\BR\left(B_u \rightarrow \tau \nu\right)}}
\newcommand*{\reftable}[1]{Table~\ref{#1}}         
       
\newcommand*{\reffig}[1]{Fig.~\ref{#1}}
 
        \newcommand*{\refeq}[1]{Eq.~(\ref{#1})}
     \newcommand*{\refsec}[1]{Sec.~\ref{#1}}

\newcommand*{\neutone}{\ensuremath{\chi^0_1}}

\let\oldcite\cite
\renewcommand*{\cite}{~\oldcite}

\newcommand*{\hl}{\ensuremath{h}}

\restylefloat{figure}

\begin{document}

\title{\LARGE {\bf Quantum gravity contributions to the gauge and Yukawa couplings in proper time flow}}
\author{\\Gabriele 
Giacometti$^{1,2}$\footnote{\email{gabriele.giacometti@phd.unict.it}},
\,Kamila Kowalska$^{3}$\footnote{\email{kamila.kowalska@ncbj.gov.pl}},
\,Daniele Rizzo$^{4}$\footnote{\email{daniele.rizzo@kbfi.ee}},\\ 
Enrico Maria Sessolo$^{3}$\footnote{\email{enrico.sessolo@ncbj.gov.pl}},
and 
Dario Zappal\`a$^{2}$\footnote{\email{dario.zappala@ct.infn.it}}\\
[2ex]
$^1$ \small{\em  
Dipartimento di Fisica e Astronomia, Universit\`a di Catania, Via S. 
Sofia 64, 95123, Catania, Italy} \\
$^2$ \small {\em INFN, Sezione di Catania, Via Santa Sofia 64, 95123 
Catania, Italy }\\
$^3$  \small {\em National Centre for Nuclear Research, Pasteura 7, 
02-093 Warsaw, Poland  }\\
$^4$ \small {\em Laboratory for High Energy and Computational 
Physics, NICPB, Rävala 10, Tallinn 10143, Estonia}
} 

%
% Date
\date{}
%\number{3}
\maketitle
\thispagestyle{fancy}

%%%%%%%%%%%%%%%%%%%%%%%%%%%%%%%%%%%%%%%%%%%%%%%%%%%%%%%%%%%%%%%%%%%%%%%%%%%%%%%%%%%%%
\begin{abstract}
We derive quantum gravity contributions to the beta functions of the gauge and Yukawa couplings of a matter theory using the Schwinger proper-time flow equation. Working in the Einstein-Hilbert truncation, we investigate the gauge-fixing and regulator dependence of the corresponding renormalization group equations. We quantify the sensitivity of our results to unphysical parameters by evaluating the gravitational correction to the running matter couplings at the interactive fixed point of gravity and we compare our findings with existing determinations in alternative schemes. 
We finally confront the derived contributions with 
the typical size they should assume to generate observable low-scale predictions in the Standard Model and in several scenarios of new physics. 
\end{abstract}
\newpage 
%%%%%%%%%%%%%%%%%%%%%%%%%%%%%%%%%%%%%%%%%

\tableofcontents

\setcounter{footnote}{0}

%%%%%%%%%%%%%%%%%%%%%%%%%%%%%%%%%%%%%%%%%%%%%%%%%%%%%%%%%%%%%%%%%%%%%%%%%%%%
\section{Introduction\label{sec:intro}}
%%%%%%%%%%%%%%%%%%%%%%%%%%%%%%%%%%%%%%%%%%%%%%%%%%%%%%%%%%%%%%%%%%%%%%%%%%%%

Asymptotic safety~(AS) is the property of a quantum field theory to develop a non-Gaussian ultraviolet~(UV) fixed point of the renormalization group~(RG) flow of the action\cite{inbookWS}. Strong indications based on the exact renormalization group~(ERG)\cite{WETTERICH199390,Morris:1993qb} point to AS being a property of quantum gravity\cite{Reuter:1996cp,Lauscher:2001ya,Reuter:2001ag,Lauscher:2002sq,Litim:2003vp,Codello:2006in,Machado:2007ea,Codello:2008vh,Benedetti:2009rx,Dietz:2012ic,Falls:2013bv,Falls:2014tra}, which can render it \textit{non-perturbatively} renormalizable. 
A profound consequence of the asymptotically safe UV behavior of quantum gravity is that a particle theory in four space-time dimensions coupled to the gravitational action may also end up being asymptotically safe if the combined system of gravity and matter develops a fixed point in the energy regime where gravitational interactions become strong\cite{Percacci:2002ie,Percacci:2003jz,Zanusso:2009bs,Daum:2009dn,Daum:2010bc,Folkerts:2011jz,Dona:2013qba,Meibohm:2015twa,Oda:2015sma,Eichhorn:2016esv,Christiansen:2017cxa,Eichhorn:2017eht,Pawlowski:2020qer,Eichhorn:2022gku}. 

We focus in this work on a renormalizable matter theory with gauge and Yukawa interactions. In the presence of a UV fixed point for quantum gravity the corresponding RG equations receive trans-Planckian contributions that, at the leading order, take the form
\bea
\frac{d g_i}{d t}&=&\beta_i^{(\textrm{matter})}-f_g\, g_i \label{eq:betag}\,,\\
\frac{d y_j}{d t}&=&\beta_j^{(\textrm{matter})}-f_y\, y_j \label{eq:betay}\,,
\eea
where $t=\ln k$ denotes the renormalization scale, $g_{i=1,2,3...}$ and $y_{j=1,2,3...}$ are, respectively, the set of gauge and Yukawa couplings, and $\beta_{i,j}^{(\textrm{matter})}$ stand for the beta functions of the matter theory without gravity. 

In Eqs.~(\ref{eq:betag}) and (\ref{eq:betay}), $f_g$ and $f_y$ indicate leading gravitational 
corrections to the running of the gauge and Yukawa couplings. They are expected to be ``universal,'' in the sense that they multiply linearly all matter couplings of the same kind, because gravity does not distinguish between internal degrees of freedom in the matter sector. As they are functions of the fixed points of the operators of the gravitational action, however, $f_g$ and $f_y$ are \textit{not} universal in the sense of being independent of computational choices like the renormalization scheme, background metric, regulator functional shape, and  gauge-fixing parameters. In the framework of the ERG, where several $f_g$ determinations, often obtained under distinct choices, exist\cite{Reuter:2001ag,Lauscher:2002sq,Percacci:2002ie,Percacci:2003jz,Codello:2007bd,Benedetti:2009rx,Narain:2009qa,Dona:2013qba,Harst:2011zx,Christiansen:2017gtg,Eichhorn:2017lry,Christiansen:2017cxa,Falls:2017lst,Falls:2018ylp,deBrito:2022vbr,Pastor-Gutierrez:2022nki,Riabokon:2025ozw}, it was established that this parameter is likely not negative, irrespective of the selected RG scheme\cite{Folkerts:2011jz,Christiansen:2017cxa}. In Ref.\cite{Folkerts:2011jz} it was shown that $f_g=0$ identically corresponds to respecting a particular classical symmetry of the gravitational action. Interestingly, however, $f_g \neq 0$ was found to arise even outside of ERG calculations, for example, in a perturbative and mass-independent scheme based on dimensional regularization\cite{Toms:2008dq,Toms:2009vd}, see also\cite{Robinson:2005fj,Pietrykowski:2006xy,Toms:2007sk,Ebert:2007gf,He:2010mt,Tang:2011gz,Felipe:2012vq,Narain:2012te,Narain:2013eea}. 
Incidentally, a large positive $f_g$ would render the entire gauge sector of the 
Standard Model~(SM) asymptotically free thus 
lending credit to the consistency of AS as a desirable UV completion. 

The gravitational contribution to the Yukawa coupling, $f_y$, was investigated in several papers in the context of the ERG\cite{Rodigast:2009zj,Zanusso:2009bs,Oda:2015sma,Eichhorn:2016esv,Hamada:2017rvn,Pastor-Gutierrez:2022nki}. While these results were somewhat inconclusive in establishing the sign of $f_y$ over the full range of possible values for the fixed point of the cosmological constant, the leading order at $\Lambda\to 0$ appears to be negative in most calculations. On the other hand, a recent study\cite{deBrito:2025nog}, which includes the complete set of operators contributing to the Yukawa-coupling beta function at next-to-leading order, seems to confidently assess that the critical exponent is large and positive when computed at the asymptotically free Yukawa-coupling fixed point, both under the assumption of a minimal matter content coupled to gravity, and in the framework of the SM, again lending consistency to the validity of AS as a viable UV completion of the SM.      

In truth, $f_g$ and $f_y$ are not observable quantities, and as such they are expected to present a certain level of gauge and scheme dependence. However, any piece of information about their size and sign obtained from first principles may lead to great progress in our understanding of some of the open issues that mar the SM at extremely high energies. This is because $f_g$ and $f_y$ are responsible for several of the very few observable signatures of asymptotically safe quantum gravity. More specifically, the \textit{a priori} free couplings of a matter Lagrangian become predictions if they correspond to \textit{irrelevant} directions of the RG flow at a UV fixed point induced by interactions with quantum gravity. Such predictions of the scale-invariant theory can in turn be tested by phenomenological means after following the RG flow of the matter couplings down to accessible energies. By means of a consistent heuristic determination of the gravitational contributions to the matter beta functions at the fixed point ($f_g$, $f_y$, as well as their counterpart for the scalar sector), it has been possible to infer a ballpark prediction for the value of the Higgs mass (more precisely, of the quartic coupling of the Higgs potential) a few years ahead of its discovery\cite{Shaposhnikov:2009pv} (see also Refs.\cite{Eichhorn:2017als,Kwapisz:2019wrl,Eichhorn:2021tsx}). A retroactive ``postdiction'' of the top-quark mass value was extracted in Ref.\cite{Eichhorn:2017ylw} and a gravity-driven prediction of the top/bottom mass ratio of the SM was obtained in Ref.\cite{Eichhorn:2018whv}. Potential imprints of UV fixed points on the flavor structure of the SM and, in particular, the Cabibbo-Kobayashi-Maskawa matrix, were found in Ref.\cite{Alkofer:2020vtb}, and equivalent analyses for the Pontecorvo-Maki-Nakagawa-Sakata matrix elements were performed in Refs.\cite{Kowalska:2022ypk} and\cite{Eichhorn:2025sux}. Predictions based on heuristic determinations of $f_g$ and $f_y$ were also extracted for several beyond-the-SM~(BSM) scenarios in relation to neutrino masses\cite{Grabowski:2018fjj,Kowalska:2022ypk,Chikkaballi:2023cce,deBrito:2025ges}, flavor anomalies\cite{Kowalska:2020gie,Chikkaballi:2022urc}, the muon anomalous magnetic moment\cite{Kowalska:2020zve}, baryon number\cite{Boos:2022jvc,Boos:2022pyq}, and dark matter\cite{Chikkaballi:2025pnw}. The asymptotically safe SM was studied in Ref.\cite{Pastor-Gutierrez:2022nki}.

Our goal in this work is to provide a determination of the parameters $f_g$ and $f_y$ independently and complementarily to the ERG. We do this by computing the Wilsonian flow of the gauge and Yukawa couplings, coupled to Einstein-Hilbert gravity, using the \textit{proper time}~(PT) 
flow equation. Rather than treating the flow of the one-particle irreducible effective 
average action, PT renormalization focuses on the flow of the Wilsonian action generated by means of the scale-dependent Schwinger PT regulator\cite{Oleszczuk:1994st,Floreanini:1995aj,Liao:1994fp,
Bohr:2000gp,deAlwis:2017ysy, Bonanno:2019ukb,
Wetterich:2024ivi}. The PT flow was previously explored in several contexts, 
from the study of renormalization properties and
determination of the scaling exponents at criticality of quantum field theories\cite{Bonanno:2004pq,
Bonanno:2000yp,Mazza:2001bp,Litim:2010tt}, to the analysis of AS in the conformal sector 
of quantum gravity\cite{Bonanno:2012dg,Bonanno:2023fij,Bonanno:2023ghc}, 
to the recent determination\cite{Bonanno:2025tfj} of the UV fixed points of the  gravitational Einstein-Hilbert action -- a determination that was shown to agree qualitatively with the ERG studies reported above. Our paper also complements a recent study that derived the gravitational corrections to the RG flow of scalar matter interactions in the PT scheme\cite{Bonanno:2026mzs}.    

We finally point out that, besides the accuracy in the determination of various observables, such as the three-dimensional Ising critical exponents, the PT flow 
is expected to preserve gauge symmetry
because of the nature of its infrared regulator\cite{Schwinger:1951nm}, 
and actually this was verified in the determination  
of the first two perturbative orders of the  
beta function of  Yang-Mills theories\cite{Liao:1995nm,Giacometti:2025qyy}.
In this sense the PT flow equation represents a convenient 
quantitative tool for the determination of quantities, such 
as $f_g$ and $f_y$, that are crucial in the study of the 
phenomenological implications of AS. 

In this paper, we extend the study in Ref.\cite{Bonanno:2025tfj} to the gauge and Yukawa couplings of the matter sector, focusing in particular on the gauge and regulator dependence of the results. Our goal is three-fold. 
Firstly, we aim at providing a quantitative counterpart to the most recent derivations of $f_g$ and $f_y$ obtained in the context of the ERG\cite{deBrito:2022vbr,Riabokon:2025ozw,deBrito:2025nog}. We shall highlight the similarities and differences that arise from our distinct choice of regularization scheme. Secondly, we provide the first, to our knowledge, determination of the Einstein-Hilbert contribution to the Yukawa beta function, $f_y$, in the PT scheme and show that it features a distinct behavior with respect to its gauge coupling counterpart. Thirdly, we compare our PT flow determinations of $f_g$ and $f_y$ with the expectations for their size that were inferred in the phenomenological studies cited above.  
 
The paper is organized as follows. In~\refsec{sec:pt} we review some of the main properties of the Schwinger PT flow equation. We employ this framework in~\refsec{sec:fgandfy} to derive quantum gravity contributions to the beta functions of the gauge and Yukawa couplings in the Einstein-Hilbert truncation. In~\refsec{sec:pheno} we study the sensitivity of our results to gauge-fixing parameters and PT regulator at the interactive UV fixed point of Einstein-Hilbert gravity, both with a minimal matter content and in the SM. We also confront our findings with phenomenological expectations on the size of gravitational corrections stemming from the requirement that the SM, and some of its well-studied extensions, remain asymptotically safe (or free) in the deep UV. We summarize our findings and present our conclusions in~\refsec{sec:con}. In Appendix~\ref{app:grav_beta} we recall the beta functions of the dimensionless counterparts of the Newton and cosmological constant in the Einstein-Hilbert truncation.

%%%%%%%%%%%%%%%%%%%%%%%%%%%%%%%%%%%%%%%%%%%%%%%%%%%%%%%%%%%%%%%%%%%%%%%%%%%%

\section{Schwinger proper time}\label{sec:pt}

In this section, following the general steps 
outlined in Ref.\cite{Bonanno:2025tfj}, 
we review some of the main properties of the 
Schwinger PT flow equation. 

We start by introducing the PT flow  equation for the 
Wilsonian  action $S_k$\cite{Oleszczuk:1994st,Floreanini:1995aj,Liao:1994fp,
Bohr:2000gp,deAlwis:2017ysy,Bonanno:2019ukb,
Wetterich:2024ivi}:
    \begin{equation}\label{eq:ptreg}
        k\partial_k S_k = -\frac{1}{2}\int_{0}^{\infty} \frac{ds}{s} \text{Tr}\left[ k \partial_k \rho_{k,k_{\textrm{UV}}}(s) e^{-s \,S_k''}\right],
    \end{equation}  
where $s$ is the PT integration variable and energy scale~$k$ 
parametrizes the RG flow.  We indicate functional derivation with ``prime'' symbols, so that $S_k''$ represents 
the second functional derivative (hessian) of the Wilsonian 
action with respect to the fields. The trace 
runs over all spatial, momentum, and internal indices of $S_k''$.  

The integral in \refeq{eq:ptreg} is regularized 
by means of  the regulating function 
$\rho_{k,k_{\textrm{UV}}}(s)$, which 
smoothly vanishes for $s> 1/k^2$ 
and $s< 1/k_{\textrm{UV}}^2$ ($k_{\textrm{UV}}$ is the UV scale at which the bare action
is defined) and is 
essentially constant in between, thus 
suppressing the integration of the field modes with 
momentum larger than $k_{\textrm{UV}}$ and 
smaller than the RG running scale~$k$.

While, at least in principle, 
all observables determined from the RG equation should  
be independent of the arbitrary details of the regulator, in practice it is convenient to select a
regulator that simplifies the computation of the integrals
and therefore, in this work, we resort to the 1-parameter 
family of regulators considered in 
Ref.\cite{Bonanno:2004sy,Bonanno:2025tfj}:
    \begin{equation}\label{eq:regm}
        \rho_{k,k_{\textrm{UV}}}(s;m,Z_k)=\frac{\Gamma(m,m\, 
        Z_k\,s\, k^2)-\Gamma(m,m\, Z_k\, s\, 
        k_{\textrm{UV}}^2)}{\Gamma(m)}\,,
    \end{equation}
where $\Gamma(m)$ indicates the gamma function and 
$\Gamma(a,x)=\int_x^{\infty}dt\,t^{a-1} e^{-t}$ denotes the incomplete gamma function. 
The parameter $m>2$, conveniently restricted to integer values, characterizes the smoothness of the shape of the regulator in proximity of the scales~$1/k^2$ and $1/k_{\textrm{UV}}^2$. In addition, we perform the rescaling $k^2\to Z_k k^2$ and $k_{\textrm{UV}}^2\to Z_k k_{\textrm{UV}}^2$ in \refeq{eq:regm} to get an effective suppression of the modes with square momentum larger than $k^2$ in the presence of a field renormalization $Z_k$\cite{Bonanno:2019ukb, Bonanno:2025tfj}.

Under these assumptions, it is straightforward to compute the derivative of $\rho_{k,k_{\textrm{UV}}}(s)$ with respect to the RG scale~$k$,
    \begin{equation}\label{eq:regulator}
        k \partial_k \rho_{k,k_{\textrm{UV}}}(s)=-\frac{2 \,
        (m\, Z_k\,s\, k^2)^m
        }{\Gamma(m)} e^{-m\,Z_k\, s\,k^2},
    \end{equation}
which we plug into \refeq{eq:ptreg} to derive the flow equation of $S_k$.

We shall also analyze the \textit{sharp limit} of the regulator in \refeq{eq:regm}, obtained for $m\to \infty$:
    \begin{equation}\label{sharpflow}
\lim_{m\to \infty} \rho_{k,k_{\textrm{UV}}}(s;m,Z_k)
= \Theta \left(\frac{1}{Z_k k^2} -s\right )-
\Theta \left(\frac{1}{Z_k k_{\textrm{UV}}^2} -s\right ),
\end{equation}
where $\Theta(x)$ is the Heaviside function. The derivative of the regulator, \refeq{eq:regulator}, becomes in this case a Dirac delta function and the PT flow equation assumes a simple form, 
    \begin{equation}\label{eq:limitflow}
        k \partial_k S_k = \text{Tr} \, e^{ -
        \frac{S_k''}{ Z_k k^2}
        }\,.
    \end{equation}
It was shown in Ref.\cite{Giacometti:2025qyy} that the sharp-limit flow in \refeq{eq:limitflow} has the quality of reproducing the two-loop beta function of Yang-Mills and scalar $O(N)$~theory, unlike the ``smooth'' regulator~(\ref{eq:regm}), which instead produces spurious $\mathcal{O}(1/m)$ terms at the two-loop order.

Once the flow equation~(\ref{eq:ptreg}) is established, 
one can derive the flow of the couplings in Eqs.~\eqref{eq:betag} and~\eqref{eq:betay} by matching 
the coefficients of the operators associated with each coupling on 
both sides of the flow equation. The traces are evaluated by using the heat kernel expansion\cite{Vassilevich:2003xt},  
        \begin{equation}
        \text{Tr} \, e^{ -s S_k''}=\frac{1}{(4\pi 
        s)^2}\sum_{n=0}^\infty\,s^n\int\, 
        d^4x\sqrt{g}\;\text{Tr}\,a_n(x)\,.
        \label{eq: Heat kernel}
    \end{equation}
For an operator of the form $S_k''\sim -\Box \mathbf{1}+\mathbb{U}$, 
we recall the explicit expression of the first few terms:   
\begin{multline}\label{eq:hk_terms}
a_0=\mathbf{1\,}, \qquad a_1=\frac{R}{6}\mathbf{1}-\mathbb{U}\,,\\
a_2=\frac{1}{2}\mathbb{U}^2+\frac{1}{6}\Box\mathbb{U}+\frac{1}{12}\mathbf{\Omega}_{\mu\nu}\mathbf{\Omega}^{\mu\nu}-\frac{1}{6}R\mathbb{U}+\mathbf{1}\left(\frac{1}{180}R_{\alpha\beta\mu\nu}^2-\frac{1}{180}R^2_{\mu\nu}+\frac{1}{72}R^2-\frac{1}{33}\Box R \right)\dots
\end{multline}
where $\Box=\nabla_{\mu}\nabla^{\mu}$, $\Omega_{\mu\nu}=[\nabla_{\mu},\nabla_{\nu}]$, and $R$, $R_{\mu\nu}$, and $R_{\alpha\beta\mu\nu}$ are the standard curvature scalar and tensors.

%%%%%%%%%%%%%%%%%%%%%%%%%%%%%%%%%%%%%%%%%%%%%%%%%%%%%%%%%%%%%%%%%%%
\section{Quantum gravity contributions to the running gauge and Yukawa couplings\label{sec:fgandfy}}
%%%%%%%%%%%%%%%%%%%%%%%%%%%%%%%%%%%%%%%%%%%%%%%%%%%%%%%%%%%%%%%%%%%%

In this section, we derive the contribution of quantum 
gravity to the beta functions of gauge and Yukawa couplings using the PT flow equation. 

%%%%%%%%%%%%%%%%%%%%%%%%%%%%%%%%%%%%%%%%%%%%%%%%%%%%
\subsection{Gauge coupling\label{sec:gauge}}

We consider an abelian gauge theory with running coupling $g_Y$. The universal nature of the gravitational interaction implies that the contribution to the beta function arising from graviton loops should be the same for any gauge coupling, independently of the group structure.   

The running gauge coupling can be extracted in different ways. One could derive the flow from the three-point function and/or the four-point function of matter/gauge boson interactions. This choice, which was explored, \textit{e.g.}, in Ref.\cite{Eichhorn:2017lry} in the context of the ERG, would require the introduction in the action of matter fields (scalars and/or fermions) charged under the abelian gauge symmetry. A more common choice\cite{Daum:2009dn,Folkerts:2011jz,Toms:2008dq,Toms:2010vy,Toms:2011zza}, which we adopt in this work,  is to extract the coupling directly from the gauge field propagator, after a rescaling of the field $A_{\mu}\to A_{\mu}/g_Y$. Of course, gauge invariance should guarantee that the beta function of $g_Y$ does not depend on the way it is extracted. In the presence of gravity, this is not obvious,  because of the perturbative non-renormalizability 
of the gravitational interaction. On the other hand, in Ref.\cite{Bevilaqua:2021uzk}
the Ward identities are shown to be respected at one- and two-loop order in quantum electrodynamics coupled to Einstein gravity, while in Ref.\cite{Souza:2022ovu} the Slavnov-Taylor identities are shown to hold at one-loop order in non-abelian, Yang-Mills theories.

Following common practice, we adopt the background field method to compute $f_g$. 
The metric tensor $g_{\mu\nu}$ 
and gauge field $A_{\mu}$ are thus split into a background and a fluctuation,
\be
g_{\mu\nu}=\bar{g}_{\mu\nu}+h_{\mu\nu}\,,\qquad A_\mu=\bar{A}_\mu+a_\mu\,.
\ee
For the rest of this subsection we shall indicate all background quantities with a bar.

We consider the following bare Wilsonian effective action, 
    \begin{align}\label{eq:action}
        S =& -\frac{1}{16\pi G} \int d^4x\, \sqrt{g} \left(R - 2\Lambda\right)+ \frac{1}{32\pi G\,\alpha}\int d^4x\sqrt{\bar{g}}\,\bar{g}^{\mu\nu}F_{\mu}F_\nu\nonumber\\
              &+ \frac{1}{4} \int d^4x\, \sqrt{g} g^{\mu\nu} g^{\kappa\lambda} F_{\mu\kappa} F_{\nu\lambda} 
              + \frac{1}{2 \xi} \int d^4x\, \sqrt{\bar{g}} \left(\bar{g}^{\mu\nu} \bar{D}_\mu a_\nu\right)^2\nonumber\\
              &+ S_{\textrm{ghost}}+S_{\textrm{matter}}\,.
    \end{align} 
The first term in \refeq{eq:action} denotes the Einstein-Hilbert action with Newton constant~$G$ and cosmological constant~$\Lambda$.  The second addend is the gauge-fixing term, where 
    \begin{equation}
        F_\mu=\left(\bar{g}^{\alpha\gamma}\delta^\beta_\mu\bar{D}_\gamma-\omega\bar{g}^{\alpha\beta}\bar{D}_\mu\right)h_{\alpha\beta}\,,
    \end{equation}
$\alpha$ and $\omega$ are gauge-fixing parameters, and the covariant derivative $\bar{D}_{\mu}$ is constructed with respect to the background metric $\bar{g}_{\mu\nu}$, 
\begin{equation}
\bar{D}_{\mu}h_{\alpha\beta}=\partial_{\mu}h_{\alpha\beta} - \bar{\Gamma}_{\mu\alpha}^{\sigma}h_{\sigma\beta}- \bar{\Gamma}_{\mu\beta}^{\sigma}h_{\alpha\sigma}\,.        
\end{equation}
In the second line of \refeq{eq:action},  $F_{\mu\nu}=\partial_\mu A_\nu-\partial_\nu A_\mu$ denotes the field strength tensor of the abelian gauge field, and $\xi$ is the gauge-fixing parameter of the abelian sector. 
Once more, the covariant derivative is defined with respect to the background metric, $\bar{D}_{\mu}a_{\nu}=\partial_{\mu}a_{\nu} - \bar{\Gamma}_{\nu\mu}^{\sigma}a_{\sigma}$. 
We do not write explicitly in \refeq{eq:action} the ghost terms relative to the gravitational and abelian gauge fixing. It is well known that ghost loops do not contribute to the running of the abelian gauge coupling, 
which is the quantity we seek to calculate here. We indicate for completeness possible contributions from the matter sector as $S_{\textrm{matter}}$ but, as was discussed above, we will not use them to extract the gravitational contribution to the gauge beta function.

We expand the action (\ref{eq:action}) to the second order in fluctuations of the metric and gauge field (gauge-fixing terms do not need to be expanded as they are second-order by construction),
\begin{equation}
\label{eq:squad}
S^{(2)}=S^{(2)}_{hh}+S^{(2)}_{ah}+S^{(2)}_{aa}\,,
\end{equation}
where
\begin{multline}
S^{(2)}_{hh}=\frac{1}{32\pi G}\int d^4x\, \sqrt{g}\left[ \frac{1}{2}\left(-h_{\mu\nu}\Box h^{\mu\nu}+h \Box h \right)+\frac{1}{2}\left(\bar{R}-2\Lambda \right) \left(h_{\mu\nu}h^{\mu\nu}-\frac{1}{2}h^2\right)\right.\\
+\left.h_{\mu\nu}\bar{D}^\mu\bar{D}^\rho h_\rho^{\,\,\nu}-h\bar{D}^\mu\bar{D}^\nu h_{\mu\nu}+h\bar{R}^{\mu\nu}h_{\mu\nu}-h_{\mu\nu}\bar{R}^{\nu\sigma}h^\mu_{\,\,\sigma}-h_{\mu\nu}\bar{R}^{\mu\rho\nu\sigma}h_{\rho\sigma}+\frac{1}{\alpha}\left(\bar{D}^{\mu} h_{\mu\nu}-\omega \bar{D}_{\nu} h \right)^2\right]\\
+\int d^4x\, \sqrt{g}\left[\frac{1}{4}\bar{F}^{\mu\nu}\bar{F}^{\rho\sigma}h_{\sigma\nu}h_{\rho\mu}+\frac{1}{2}\bar{F}^{\mu\nu}\bar{F}_{\mu}^{\,\,\rho}\left(h_\nu^{\,\,\sigma}h_{\rho\sigma}-\frac{1}{2}h h_{\nu\rho}\right)
-\frac{1}{16}\bar{F}_{\rho\sigma}\bar{F}^{\rho\sigma}\left(h_{\mu\nu}h^{\mu\nu}-\frac{1}{2}h^2\right)\right],
\label{eq:squadhh}
\end{multline}
\begin{equation}
S^{(2)}_{ah}=-\frac{1}{2}\int d^4x\, \sqrt{g}\left(h \bar{F}_{\nu\mu}-2h_{\mu\rho}\bar{F}_\nu^{\,\,\rho}+2h_{\nu\rho}\bar{F}_\mu^{\,\,\rho}\right)\bar{D}^\mu a^\nu\,,
\label{eq:squadah}
\end{equation}
\begin{equation}\label{eq:s2}
S^{(2)}_{aa}=\int d^4x\, \sqrt{g}\left[\frac{1}{2\xi}(\bar{D}_\mu a^\mu)^2+\frac{1}{2}\left(-a_\mu\Box a^\mu+a_\mu \bar{D}^\mu\bar{D}^\nu a_\nu\right)\right].
\end{equation}
In Eqs.~(\ref{eq:squadhh})-(\ref{eq:s2}), the curvature tensors are constructed with respect to the background metric $\bar{g}_{\mu\nu}$, we have defined $h\equiv h_{\mu}^{\mu}$ and we have introduced the d'Alambertian, $\Box=\bar{D}^\mu\bar{D}_\mu$.

The action in \refeq{eq:squad} can be recast into a $2\times 2$ hessian block matrix, following the procedure delineated, \textit{e.g.}, in Ref.\cite{Codello:2015oqa}. One writes
\be
S^{(2)}=\frac{1}{32\pi G}\int d^4 x \sqrt{g} 
\left(h_{\alpha\beta},a_{\rho}\right) \left( {\begin{array}{cc}
\mathbb{H}_{hh}^{\alpha\beta\mu\nu} & \mathbb{H}_{ha}^{\alpha\beta\sigma}\\
\mathbb{H}_{ah}^{\rho\mu\nu} & \mathbb{H}_{aa}^{\rho\sigma}
 \end{array} } \right) \left( {\begin{array}{c}
h_{\mu\nu}\\
a_{\sigma}
 \end{array} } \right),
\ee
where matrices $\mathbb{H}$ take the form 
\be\label{eq:hes_mat}
\mathbb{H}=-\mathbb{K} \Box+2 \mathbb{V}^{\delta}\bar{D}_{\delta}+\mathbb{U}\,,
\ee
and matrices $\mathbb{K}$, $\mathbb{V}^{\delta}$, and potential $\mathbb{U}$ can be extracted, with some work, from Eqs.~(\ref{eq:squadhh})-(\ref{eq:s2}). Tensor manipulations in this paper are performed with the help of the \texttt{xAct}\cite{Brizuela:2008ra,Martin-Garcia:2007bqa,Martin-Garcia:2008yei,Martin-Garcia:2008ysv} Mathematica package.

We perform the calculation in the de Donder gauge, $\omega=1/2$. Moreover, we adopt a flat Euclidean background, which simplifies the inversion of operators enormously. With this choice, one can set 
\be\label{eq:assume}
\bar{g}_{\mu\nu}=\delta_{\mu\nu}\,,\quad \quad \bar{R}=0\,,\quad \quad 
\bar{D}_{\mu}\bar{D}_{\nu}=-p_{\mu} p_{\nu}\,.
\ee
Our background choice implies that the calculation is by necessity off-shell, since the flat metric is not a solution to the Einstein equations of motion with nonzero cosmological constant. A direct consequence of this observation is that our result for the beta function will be dependent on the gauge fixing of the gravitational sector\cite{Toms:2007sk,Toms:2008dq}. 

Under assumptions~(\ref{eq:assume}), and after rescaling the photon field $a_{\mu}\to \sqrt{16\pi G} \, a_{\mu}$, it is possible to bring \refeq{eq:hes_mat} to its canonical form in a general covariant gauge by contracting it with the inverse operators\cite{Toms:2011zza},
\be\label{eq:propgr}
\mathbb{P}^{\alpha\beta\mu\nu}_{hh}=g^{\alpha\mu}g^{\beta\nu}+g^{\alpha\nu}g^{\beta\mu}-g^{\alpha\beta}g^{\mu\nu}
-\left(1-\alpha\right)
\frac{g^{\alpha\mu}p^{\beta}p^{\nu}+g^{\alpha\nu}p^{\beta}p^{\mu}+g^{\beta\mu}p^{\alpha}p^{\nu}+g^{\beta\nu}p^{\alpha}p^{\mu}}{p^2}\,,
\ee
\be\label{eq:propga}
\mathbb{P}^{\rho\sigma}_{aa}=\left(16\pi G \right)^{-1}\left(g^{\rho\sigma}-\left(1-\xi\right)\frac{p^{\rho} p^{\sigma}}{p^2}\right)\,.
\ee
As a consequence, potential $\mathbb{U}$ undergoes a rescaling and its diagonal blocks become
\bea
\mathbb{W}^{\alpha\beta\mu\nu}_{hh}&=&\mathbb{P}_{hh\,\lam\kappa}^{\alpha\beta}\mathbb{U}_{hh}^{\lam\kappa\mu\nu},\\
\mathbb{W}^{\rho\sigma}_{aa}&=&\mathbb{P}_{aa\,\kappa}^{\rho}\mathbb{U}_{aa}^{\kappa\sigma}\,.\label{eq:rephot}
\eea

Following standard procedure, we next absorb the linear $\mathbb{V}^{\delta}\bar{D}_{\delta}$ term into a 
redefinition of the covariant derivative, as first prescribed in Ref.\cite{Jackiw:1978ar}. One can define 
\bea
\mathbb{Y}_{ha}^{\alpha\beta\delta\sigma}&=&\mathbb{P}_{hh\,\lam\kappa}^{\alpha\beta}\mathbb{V}_{ha}^{\lam\kappa\delta\sigma},\\
\mathbb{Y}_{ah}^{\rho\delta\mu\nu}&=&\mathbb{P}_{aa\,\kappa}^{\rho}\mathbb{V}_{ah}^{\kappa\delta\mu\nu}.
\eea
As a result, one obtains a new Laplacian operator and a canonical hessian,
\be
\mathbb{\tilde{H}}=-\Box+\mathbb{\tilde{W}}\,.
\ee
The new potential, $\mathbb{\tilde{W}}$, is block diagonal in field space and takes the form
\bea
\mathbb{\tilde{W}}_{hh}^{\alpha\beta\mu\nu}&=&\mathbb{W}_{hh}^{\alpha\beta\mu\nu}+\mathbb{Y}_{ha}^{\alpha\beta\lam\kappa}\mathbb{Y}_{ah\,\kappa}^{\,\tau\mu\nu}\left(\frac{4 p_{\lam} p_{\tau}}{p^2}\right),\label{eq:Wtilgrav}\\
\mathbb{\tilde{W}}^{\rho\sigma}_{aa}&=&\mathbb{W}^{\rho\sigma}_{aa}+\mathbb{Y}_{ah}^{\rho\lam\mu\nu}\mathbb{Y}_{ha\,\mu\nu}^{\,\,\tau\sigma}\left(\frac{4 p_{\lam} p_{\tau}}{p^2}\right),\label{eq:Wtilphot}
\eea
where, in the language of perturbation theory, the trace of the first addend on the r.h.s.~of \refeq{eq:Wtilgrav} provides the ``tadpole'' diagram contribution to the background $\bar{F}_{\mu\nu}\bar{F}^{\mu\nu}$ operator, whereas the traces of the second addends~in Eqs.~(\ref{eq:Wtilgrav}), (\ref{eq:Wtilphot}) correspond to the ``sunset'' diagrams. 

Finally, we compute the heat kernel expansion in \refeq{eq: Heat kernel}
by making the identification $S_k''\equiv \mathbb{\tilde{H}}$. 
The first order in the expansion reads
\be\label{eq:gau_or1}
\textrm{Tr}\,a_1\equiv  -\textrm{Tr}\,\mathbb{\tilde{W}}=
12\Lambda+8\alpha\Lambda+12\pi G\left(1+\alpha\right)\bar{F}_{\mu\nu}\bar{F}^{\mu\nu}\,.
\ee
Note that the gauge parameter~$\xi$ cancels out of \refeq{eq:gau_or1}, which implies that the result is gauge-fixing independent in the abelian sector. Incidentally, \refeq{eq:gau_or1} agrees with previous determinations in the same scheme\cite{Toms:2010vy,Toms:2011zza}.

The terms containing the 
cosmological constant in \refeq{eq:gau_or1} emerge from the trace of the $\mathbb{\tilde{W}}_{hh}^{\alpha\beta\mu\nu}$ block~(\ref{eq:Wtilgrav}). Conversely, the  coefficient left to multiply the field strength tensor operator $\bar{F}_{\mu\nu}\bar{F}^{\mu\nu}$ arises from the trace of the $\mathbb{\tilde{W}}^{\rho\sigma}_{aa}$ block~(\ref{eq:Wtilphot}). The contributions from tadpole and sunset in \refeq{eq:Wtilgrav}, both in size equal to $12\pi G(1+\alpha)$, cancel out. As a consequence, due to the block-diagonal nature of the trace, we do not expect the gravitational contribution to the photon two-point function to depend on the cosmological constant at any order in the heat kernel expansion. 
This is confirmed with a direct calculation, which shows that
\be\label{eq:ngrt1}
\textrm{Tr}\,a_n|_{\bar{F}_{\mu\nu}\bar{F}^{\mu\nu}}\equiv \frac{1}{n!} \textrm{Tr}\,\mathbb{\tilde{W}}^n|_{\bar{F}_{\mu\nu}\bar{F}^{\mu\nu}} =0\,, \qquad \textrm{for }n>1\,.
\ee

We perform the integral on the r.h.s.~of \refeq{eq:ptreg} and extract the beta function from the anomalous dimension of the two-point function on the l.h.s. Defining $\eta_A\equiv -k\partial_k Z_{A,k}$ and $\beta_{g_Y}=1/2\, \eta_A g_Y$, one computes
\be
k\partial_k S_k|_{\bar{F}_{\mu\nu}\bar{F}^{\mu\nu}} \equiv k\partial_k \left(\frac{Z_{A,k}}{4}\int d^4 x \bar{F}_{\mu\nu} \bar{F}^{\mu\nu}\right)=
-\frac{1}{2}\int_{0}^{\infty} \frac{ds}{s} \text{Tr}\left[ k \partial_k \rho_{k,k_{\textrm{UV}}}(s)\, e^{-s \,S_k''}\right]|_{\bar{F}_{\mu\nu}\bar{F}^{\mu\nu}}.
\ee

The regulator-dependent gravitational contribution to the beta function~(\ref{eq:betag}) finally reads
\be\label{eq:fgfinal}
f_g=\frac{3 m \tilde{G}}{2\pi\left( m-1\right)}\left(1+\alpha\right),
\ee
where $\tilde{G}=G k^2$ is the dimensionless Newton constant. In the limit of sharp regulator, $m\to \infty$, one finds 
\be\label{eq:fgsharp}
f_g=\frac{3 \tilde{G}}{2 \pi} \left(1+\alpha\right).
\ee
The result is positive for $\alpha>-1$ and shows remarkable stability under our regulator choice for $m\gsim 20$, where it essentially comes to coincide with the sharp regulator limit. For $m=10$ the difference to the sharp limit is about 10\% and for $m=20$ it is 5\%.

We note that $f_g$ arises from the quadratic divergence element of the heat kernel expansion, $\textrm{Tr}\,a_1$ in \refeq{eq:gau_or1}, 
and it is thus proportional to the trans-Planckian fixed-point value of $\tilde{G}$.
This is in agreement with the leading term of known ERG calculations at $\alpha=\xi=1$\cite{Daum:2009dn,Folkerts:2011jz}, which differs from \refeq{eq:fgfinal} in the form of the regulator only. Equation~(\ref{eq:fgfinal}) is independent of the cosmological constant. However -- as was discussed, \textit{e.g.}, in Ref.\cite{Daum:2009dn} -- the dependence on the cosmological constant can be reintroduced via renormalization-group improvement of the calculation, \textit{i.e.}, by including the scale derivative of $Z_k$ and the anomalous dimension of the graviton in \refeq{eq:regm}. 

The next-to-leading order in the heat kernel expansion, $\textrm{Tr}\,a_2$, may include in principle a 
$G\Lambda \bar{F}_{\mu\nu}\bar{F}^{\mu\nu}$ term, corresponding to the logarithmic divergence. However, in flat background, expansion orders beyond the first do not give corrections to the $\bar{F}_{\mu\nu}\bar{F}^{\mu\nu}$ operator. This is a consequence of the Ward identity: the photon remains strictly massless in the action~(\ref{eq:squadhh})-(\ref{eq:s2}), unlike the graviton, which acquires instead an effective mass proportional to the cosmological constant.
It was shown in Refs.\cite{Toms:2008dq,Toms:2009vd}, that the logarithmic divergence can be computed in the (gauge-fixing independent) framework of the Vilkovisky-DeWitt geometrical action\cite{Vilkovisky:1984st,Vilkovisky1984qtg,DeWitt1987}, that is to say, 
by adding to Eqs.~(\ref{eq:squadhh})-(\ref{eq:s2}) terms originating 
from a metric connection applied to field space, which end up generating an effective mass for the photon. 

Despite the fact that our $f_g$ results are, quite expectedly, gauge-fixing choice dependent, it is important to be aware that this is the usual situation with beta functions of marginal couplings that depend on dimensionful coefficients like Newton and the cosmological constant. To this regard, we reiterate that the couplings of the effective action are not observable physical quantities -- which should be completely independent of the choices made for computational convenience -- only the critical exponents are physical. We expect that the gauge dependence of the beta function will be, at least in part, absorbed in computations of observable quantities like, \textit{e.g.}, the $S$-matrix. On the other hand, some residual gauge and regulator dependence may also be possibly interpreted as indication of an insufficient truncation in the action.      

%%%%%%%%%%%%%%%%%%%%%%%%%%%%%%%%%%%%%%%%%%%%%%%%%%%%%%%%%%%%%%%%%%%
\subsection{Yukawa coupling\label{sec:Yukawa}}

We now compute the contribution of minimally coupled Einstein-Hilbert gravity to the running of the Yukawa-coupling operator. We add the following term to the bare 
effective action:
\be\label{eq:Yuk_op}
S_{\textrm{Yuk}}= \int d^4x \, \sqrt{g} \, y\,\bar{\psi}\phi\psi\,,
\ee
where $\phi$ is a scalar field, which we choose to be real without loss of generality, and $\psi$ is a Dirac fermion. We limit ourselves to computing the graviton tadpole corrections to \refeq{eq:Yuk_op}, which provides the dominant term at $|\tilde{\Lambda}|\ll 1$ and allows for a straightforward resummation of heat kernel terms to all orders in $\tilde{\Lambda}$. 
As before, we work in the background field method,
\be
g_{\mu\nu}=\bar{g}_{\mu\nu}+h_{\mu\nu}\,,\qquad \psi=\tilde{\psi}+\delta\psi\,,\qquad \phi=\tilde{\phi}+\delta\phi\,,
\ee
with flat background metric $\bar{g}_{\mu\nu}=\delta_{\mu\nu}$ and arbitrary background matter fields
$\tilde{\psi}$ and $\tilde{\phi}$ (note that we indicate the background fields in the Yukawa operator with a tilde to avoid confusion with the Dirac conjugation of the fermions).

After an expansion in the fluctuating fields, the hessian takes a $3 \times 3$ matrix form,
\be
S^{(2)}=\frac{1}{32\pi G}\int d^4 x \sqrt{g} 
\left(h_{\alpha\beta},\delta\psi^{\dag},\delta\phi\right) \left( {\begin{array}{ccc}
\mathbb{H}_{hh}^{\alpha\beta\mu\nu} & \mathbb{H}_{h\psi}^{\alpha\beta} & \mathbb{H}_{h\phi}^{\alpha\beta} \\
\mathbb{H}_{\bar{\psi} h}^{\mu\nu} & \mathbb{H}_{\bar{\psi}\psi} & \mathbb{H}_{\bar{\psi}\phi} \\
\mathbb{H}_{\phi h}^{\mu\nu} & \mathbb{H}_{\phi\psi} & \mathbb{H}_{\phi\phi} 
 \end{array} } \right) \left( {\begin{array}{c}
h_{\mu\nu}\\
\delta\psi\\
\delta\phi
 \end{array} } \right).
\ee
The off-diagonal term mixing gravity and matter, extracted from \refeq{eq:Yuk_op}, reads
\be\label{eq:mix_yuk}
S^{(2)}_{\textrm{mix}}=\int d^4 x \sqrt{g}\, \frac{y}{2}\, h \left(\delta\psi^{\dag}\tilde{\phi}\tilde{\psi}+ \bar{\tilde{\psi}}\,\delta\phi\, \tilde{\psi}+ \bar{\tilde{\psi}}\tilde{\phi}\,\delta\psi \right).
\ee
It is easy to see that no graviton loop originating from \refeq{eq:mix_yuk} can provide a contribution to the renormalization of the background Yukawa operator, $y\bar{\tilde{\psi}}\tilde{\phi}\,\tilde{\psi}$, at the leading order in $G$ of the heat kernel expansion. Thus, we can safely focus on the contributions belonging to the $\mathbb{H}_{hh}^{\alpha\beta\mu\nu}$ term, whose expansion was given in \refeq{eq:squadhh}.

The hessian takes then the form
\be
    \mathbb{H} = -\Box + \mathbb{W}\,,
\ee
where 
\be
    \mathbb{W}^{\alpha\beta\mu\nu}=\left(\Lambda+8\pi G y\,\bar{\tilde{\psi}}\tilde{\phi}\,\tilde{\psi}\right)\mathcal{O}^{\alpha\beta\mu\nu}
\ee
and
\be
\mathcal{O}^{\alpha\beta\mu\nu}=-\left(g^{\alpha\mu} g^{\beta\nu}+g^{\alpha\nu} g^{\beta\mu}\right)+\frac{\alpha-1}{p^2}\left[2 g^{\mu\nu} p^{\alpha} p^{\beta}-\left( g^{\beta\nu} p^{\alpha} p^{\mu}+g^{\alpha\nu} p^{\beta} p^{\mu}+g^{\beta\mu} p^{\alpha} p^{\nu}+g^{\alpha\mu} p^{\beta} p^{\nu} \right)\right].
\ee

In this case, \refeq{eq: Heat kernel} reduces to 
\be
\textrm{Tr}\, e^{-s S_k''}=\frac{1}{(4\pi s)^2}\int d^4 x \sqrt{g}\, \textrm{Tr}\, e^{-s\left(\Lambda+8\pi G y\,\bar{\tilde{\psi}}\tilde{\phi}\,\tilde{\psi} \right)\mathcal{O}}.
\ee
The operator $\mathcal{O}$ can 
be decomposed as
\be 
    \mathcal{O}=
    -2\left(\mathbb{I}-\mathbb{X}\right) -2\alpha\,\mathbb{X}= -2\left[\mathbb{I}+(\alpha-1)\mathbb{X}\right].
\ee
Here $\mathbb{I}$ is the identity on symmetric rank-two tensors,
\begin{equation}
\mathbb{I}^{\alpha\beta}_{\mu\nu} = \frac12 \left(\delta^\alpha_\mu \delta^\beta_\nu + \delta^\alpha_\nu \delta^\beta_\mu \right),
\end{equation}
and
\begin{equation}
\mathbb{X}^{\alpha\beta}_{\mu\nu} = \frac{1}{2p^2}
\left( \delta^\beta_\nu p^\alpha p_\mu + \delta^\alpha_\nu p^\beta p_\mu + \delta^\beta_\mu p^\alpha p_\nu + \delta^\alpha_\mu p^\beta p_\nu - 2 g_{\mu\nu} p^\alpha p^\beta \right).
\end{equation}

With this definition one has
\begin{equation}
\mathbb{X}^2 = \mathbb{X}, \qquad \qquad \left(\mathbb{I}-\mathbb{X}\right)^2 = \mathbb{I}-\mathbb{X}, \qquad \qquad \left(\mathbb{I}-\mathbb{X}\right)\mathbb{X}=0\,.
\end{equation}
Therefore, $\mathbb{X}$ and $\mathbb{I}-\mathbb{X}$ are complementary projectors. Notice that these are not the usual trace and traceless projectors. Rather, $\mathbb{X}$ projects onto the momentum-longitudinal sector associated with $p_\mu$, while $\mathbb{I}-\mathbb{X}$ projects onto its complementary subspace. In $d=4$, the dimension of such subspaces are
respectively 6 and 4. 
Using the decomposition above, the trace of the exponential of the operator $\mathcal{O}$ is given by the sum of the exponentials of the two eigenvalues, multiplied by their corresponding multiplicities:
\begin{equation}\label{eq:resum} 
    \mathrm{Tr}\, e^{-s \left(\Lambda+8\pi G y\,\bar{\tilde{\psi}}\tilde{\phi}\tilde{\psi}\right)\mathcal{O}} = 6 \, e^{2s\Lambda} e^{2s\left(8\pi G y\, \bar{\tilde{\psi}} \tilde{\phi} \tilde{\psi} \right)} + 4 \, e^{2\alpha s\Lambda} e^{2\alpha s\left(8\pi G y \, \bar{\tilde{\psi}} \tilde{\phi} \tilde{\psi}\right)}\,,
\end{equation}
where we have used the fact that $\Lambda$ is a constant with respect to both momenta and fields and therefore commutes with both of them.

After expanding \refeq{eq:resum} we retain the linear order in the Yukawa operator and perform the heat kernel integral, thus finally obtaining a closed-form expression for the tadpole contribution to the Yukawa beta function~(\ref{eq:betay}) valid to all orders in $\tilde{\Lambda}=\Lambda k^{-2}$,
\be\label{eq:fyfinal}
f_y=-  \frac{2\tilde{G}}{\pi}\frac{m}{m-1}\left[3\left(1-2\frac{\tilde{\Lambda}}{m} \right)^{1-m}+2\alpha\left(1-2\alpha\frac{\tilde{\Lambda}}{m} \right)^{1-m} \right].
\ee
Taking the sharp-regulator limit, $m\to \infty$, one gets,
\be\label{eq:fysharp}
f_y=- \frac{2\tilde{G}}{\pi} \left(3\, e^{2\tilde{\Lambda}}+2\alpha\, e^{2\alpha\tilde{\Lambda}} \right).
\ee

%%%%%%%%%%%%%%%%%%%%%%%%%%%%%%%%%%%%%%%%%%%%%%%%%%%%%%%%%%%%%%%%%%%%%%%
 \begin{figure}[t]
	\centering%
  		\includegraphics[width=0.45\textwidth]{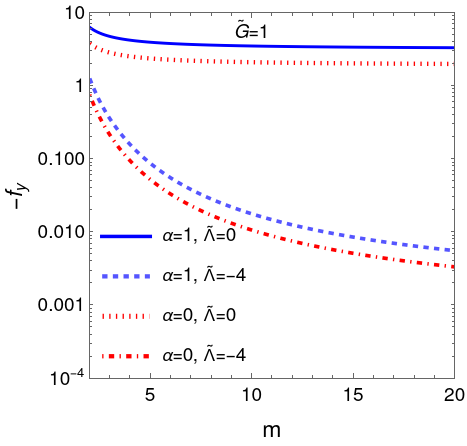}\\
\caption{$-f_y$ as a function of the regulator $m$ for $\tilde{G}=1$. Solid blue: $\alpha=1$, $\tilde{\Lambda}=0$; dashed blue: $\alpha=1$, $\tilde{\Lambda}=-4$; dotted red: $\alpha=0$, $\tilde{\Lambda}=0$; dot-dashed red: $\alpha=0$, $\tilde{\Lambda}=-4$.}
\label{fig:plot_params}
\end{figure}
%%%%%%%%%%%%%%%%%%%%%%%%%%%%%%%%%%%%%%%%%%%%%%%%%%%%%%%%%%%%%%%%%%%%%

Equations~(\ref{eq:fyfinal}), (\ref{eq:fysharp}) are negative for $\alpha>0$. In \reffig{fig:plot_params} we show the regulator dependence of~$-f_y$ at the baseline value of Newton constant, $\tilde{G}=1$, for different values of the cosmological constant. Solid and dashed blue lines show the Feynman gauge, $\alpha=1$, for $\tilde{\Lambda}=0$ and $\tilde{\Lambda}=-4$, respectively. Corresponding $f_y$ values in the Landau gauge, $\alpha=0$, 
are given by the dotted and dot-dashed red lines, respectively. 
Negative values of the cosmological constant induce an exponential suppression of the gravitational contribution to the Yukawa coupling, which, in the sharp regulator limit, can thus become comparable in size to the typical anomalous dimensions of renormalizable matter couplings in the absence of gravity. We will analyze the phenomenological consequences of this suppression in \refsec{sec:pheno}.

%%%%%%%%%%%%%%%%%%%%%%%%%%%%%%%%%%%%%%%%%%%%%%%%%%%%%%%%%%%%%%%%%%%%%%%
\section{Expectations for $\boldsymbol{f_g}$ and $\boldsymbol{f_y}$ in particle physics\label{sec:pheno}}
%%%%%%%%%%%%%%%%%%%%%%%%%%%%%%%%%%%%%%%%%%%%%%%%%%%%%%%%%%%%%%%%%%%%%%

In this section, we evaluate the values of $f_g$ and $f_y$ at the interactive UV~fixed point of the Einstein-Hilbert action, $(\tilde{\Lambda}^{\ast},\tilde{G}^{\ast})\neq (0,0)$, which was recently computed in the PT flow analysis of Ref.\cite{Bonanno:2025tfj}. We take the gauge and regulator dependence of the $\tilde{G}$ and $\tilde{\Lambda}$ beta functions at face value, neglecting in a first approximation the fact that they were obtained in a pure gravity setting and thus do not include the effects of the abelian gauge ghosts, which might induce some corrections.   
Additionally, the beta functions of the gravitational couplings are modified by the inclusion of 
matter fields, in similar fashion to the beta functions derived in the context of the ERG, \textit{e.g.}, in Ref.\cite{Dona:2013qba}. In light of the universality of gravitational interactions, the matter content is treated generically by adding an appropriate number of real scalar 
($N_S$), Dirac fermion ($N_D$), and vector ($N_V$) fields. We report for completeness, in Appendix~\ref{app:grav_beta}, the beta functions of the gravitational sector computed in Ref.\cite{Bonanno:2025tfj}. Their regulator dependence at selected $\alpha$-gauge fixing and matter content is shown in \reffig{fig:gravitym} of Appendix~\ref{app:grav_beta}.

%%%%%%%%%%%%%%%%%%%%%%%%%%%%%%%%%%%%%%%%%%%%%%%%%%%%%%%%%%%%%%%%%%%%%%%%%%%%
\subsection{Minimal matter content\label{sec:min_mat}}

We start our discussion by showing the parametric dependence of $f_g$ and $f_y$ in a model with minimal matter content, $N_S=N_D=N_V=1$, to allow for direct comparison of our PT scheme with recent ERG-based studies with the same matter content\cite{deBrito:2022vbr,Riabokon:2025ozw}. 

%%%%%%%%%%%%%%%%%%%%%%%%%%%%%%%%%%%%%%%%%%%%%%%%%%%%%%%%%%%%%%%%%%%%%%%%%%%%%%%%
\begin{figure}[t]
	\centering%
 	    \subfloat[]{%
		\includegraphics[width=0.48\textwidth]
        {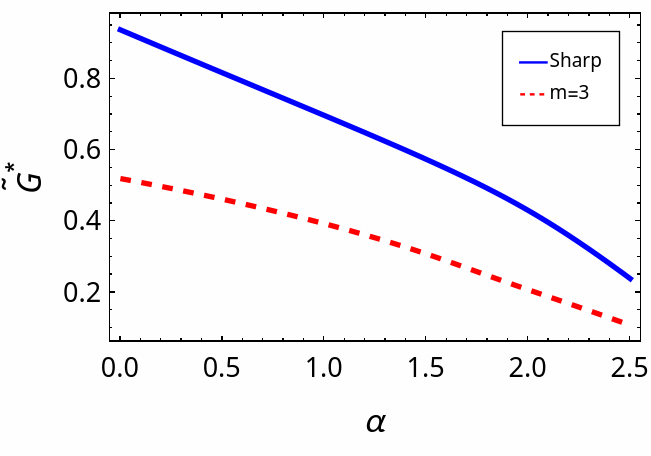}}
  		\hspace{0.2cm}
    	\subfloat[]{%
  		\includegraphics[width=0.48\textwidth]
        {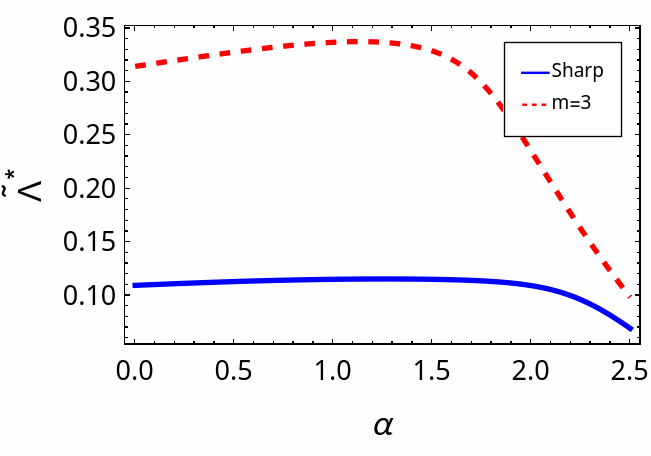}}\\
 	    \subfloat[]{%
		\includegraphics[width=0.48\textwidth]
       {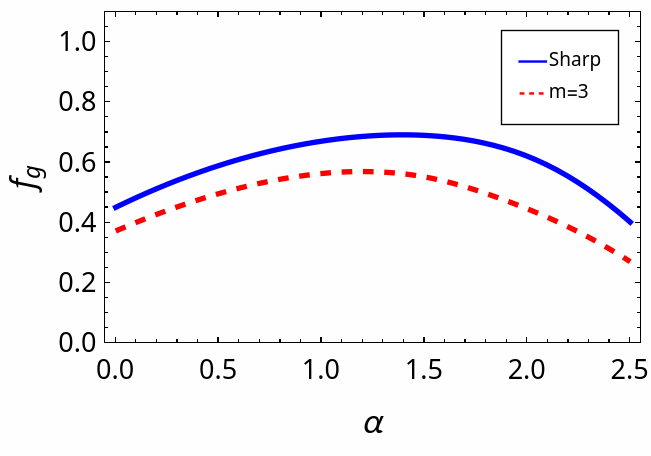}}
  		\hspace{0.2cm}
    	\subfloat[]{%
  		\includegraphics[width=0.48\textwidth]
        {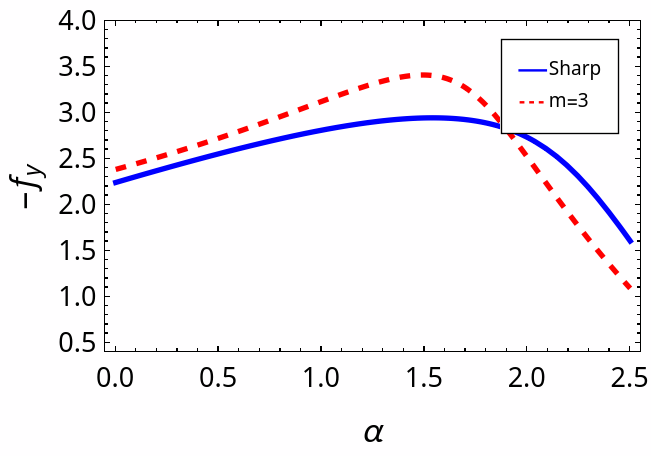}}\\
\caption{Dependence on the gauge-fixing parameter $\alpha$ of the gravity fixed point (a)~$\tilde{G}^\ast$ and (b)~$\tilde{\Lambda}^\ast$ in the minimal matter model. Dependence on the gauge-fixing parameter $\alpha$ of (c)~$f_g$ and (d)~$-f_y$ in the minimal matter model. Blue solid line indicates the sharp regulator ($m\to \infty$), while dashed red line corresponds to a specific choice of the regulator parameter, $m=3$.}
\label{fig:minimal_gauge}
\end{figure}
%%%%%%%%%%%%%%%%%%%%%%%%%%%%%%%%%%%%%%%%%%%%%%%%%%%%%%%%%%%%%%%%%%%%%%%%%%%%%%%%%%%%%

The dependence on the gauge-fixing parameter~$\alpha$ of the gravity fixed point $\tilde{G}^\ast$ 
is recalled for convenience in \reffig{fig:minimal_gauge}(a) and the dependence of $\tilde{\Lambda}^\ast$ is presented in \reffig{fig:minimal_gauge}(b). As can be seen in \reffig{fig:gravitym} of Appendix~\ref{app:grav_beta}, $\tilde{G}^\ast$ and $\tilde{\Lambda}^\ast$  are monotonic functions of the regulator, so that it is sufficient to show two benchmarks values: $m=3$~(dashed red) and $m\to\infty$~(solid blue). 

In \reffig{fig:minimal_gauge}(c) we show the dependence of $f_g$ on the gauge-fixing parameter~$\alpha$ for the same two choices of regulator~$m$.  The size of $f_g$ shows relative stability under variations of~$m$, with fluctuations of approximately~$25\%$ around the average, and also under gauge-fixing choices, which induce size variations that do not exceed an overall~$75\%$. We also find remarkable agreement with recent ERG studies\cite{deBrito:2022vbr,Riabokon:2025ozw}, which computed $f_g$ at $\alpha=\xi=0$. It gives yet another encouraging indication that the result in the gauge sector is relatively independent of the selected renormalization scheme. 

The dependence of $f_y$ on~$\alpha$ is shown in \reffig{fig:minimal_gauge}(d). $f_y$ is negative and, similarly to $f_g$, shows relative stability under the choice of regulator and gauge-fixing parameter. We observe approximately~$25\%$ swings above and below the average value of $f_y$ over the full $\alpha$ range. As one can evince from \refeq{eq:fyfinal}, the fixed-point value of the cosmological constant can have a significant impact on the size of $f_y$. On the other hand, as \reffig{fig:minimal_gauge}(b) shows, in minimal matter $\tilde{\Lambda}^{\ast}>0$. As a consequence, $f_y$ receives an enhancement and settles at relatively large negative values. 

%%%%%%%%%%%%%%%%%%%%%%%%%%%%%%%%%%%%%%%%%%%%%%%%%%%%%%%%%%%%%%%%%%%%%%%%%%%%
\subsection{Standard Model\label{sec:SM_mat}}

%%%%%%%%%%%%%%%%%%%%%%%%%%%%%%%%%%%%%%%%%%%%%%%%%%%%%%%%%%%%%%%%%%%%%%%%%%%%%%%%%%%%%%%%
\begin{figure}[t]
	\centering%
 	    \subfloat[]{%
		\includegraphics[width=0.48\textwidth]
        {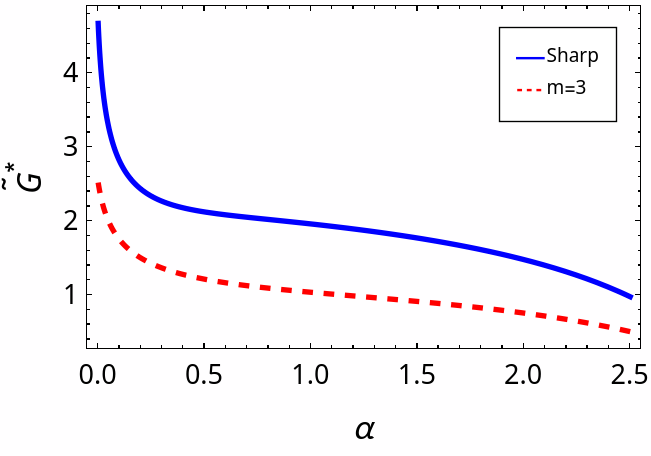}}
  		\hspace{0.2cm}
    	\subfloat[]{%
  		\includegraphics[width=0.48\textwidth]
        {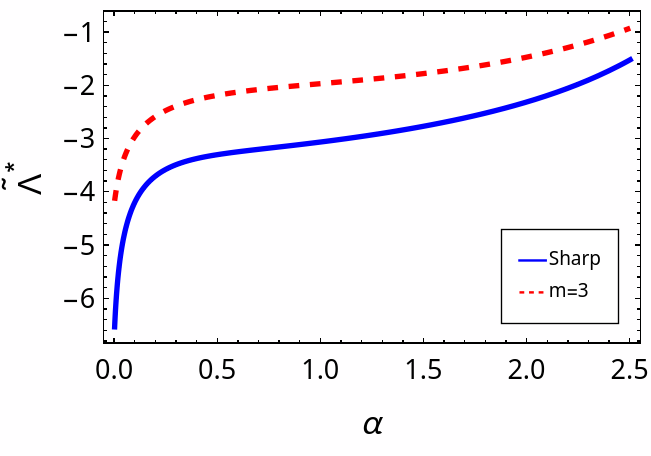}}\\
       	    \subfloat[]{%
		\includegraphics[width=0.48\textwidth]
        {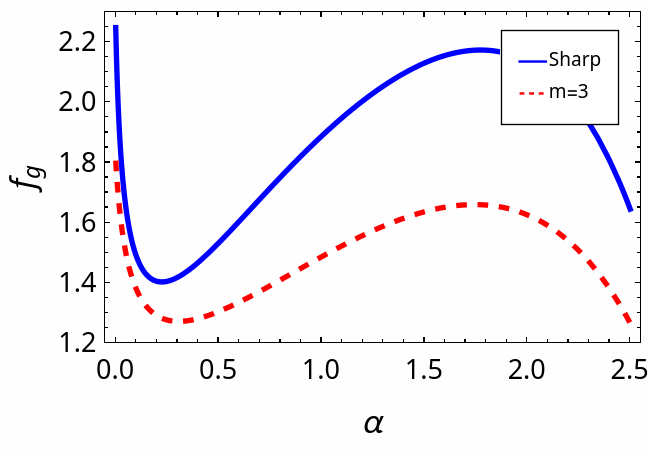}}
  		\hspace{0.2cm}
    	\subfloat[]{%
  		\includegraphics[width=0.48\textwidth]
        {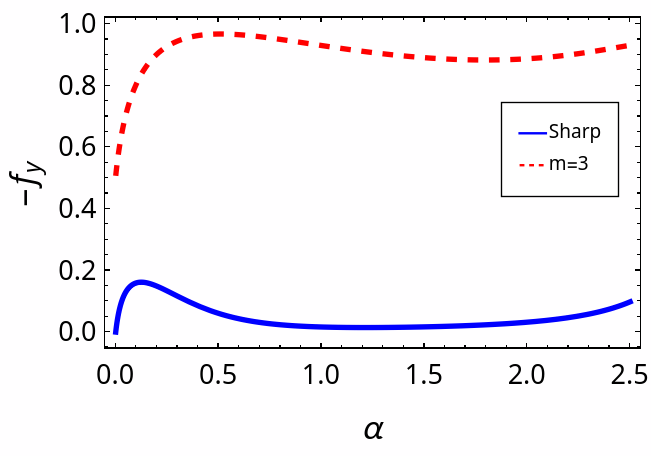}}\\  
\caption{Dependence on the gauge-fixing parameter $\alpha$ of the gravity fixed point (a)~$\tilde{G}^\ast$ and (b)~$\tilde{\Lambda}^\ast$ in the SM. Dependence on the gauge-fixing parameter $\alpha$ of (c)~$f_g$ and (d)~$-f_y$ in the SM. Blue solid line indicates the sharp regulator ($m\to \infty$), while dashed red line corresponds to a specific choice of the regulator parameter, $m=3$.}
\label{fig:SM_gauge}
\end{figure}
%%%%%%%%%%%%%%%%%%%%%%%%%%%%%%%%%%%%%%%%%%%%%%%%%%%%%%%%%%%%%%%%%%%%%%%%%%%%%%%%%%%%%%%%%%%%

We next proceed to analyze the parametric dependence of $f_g$ and $f_y$ in the SM: $N_S=4$, $N_D=45/2$,  $N_V=12$. As before, we recall the dependence on the gauge-fixing parameter~$\alpha$ 
of the gravity fixed point $\tilde{G}^\ast$ in \reffig{fig:SM_gauge}(a) and the $\alpha$-dependence of $\tilde{\Lambda}^\ast$ in \reffig{fig:SM_gauge}(b), at two benchmark values of~$m$. We have checked that across the full $\alpha$ spectrum, the anomalous dimension of the graviton, $\eta_G$ in \refeq{eq:etagrav} of Appendix~\ref{app:grav_beta}, remains firmly around~$-2$. 

The $\alpha$-dependence of $f_g$ is shown in \reffig{fig:SM_gauge}(c) for $m=3$~(dashed red) and $m\to\infty$~(solid blue). Unlike the case of minimal matter, the result appears to be highly sensitive to the gauge-fixing choice. We observe a rollercoaster-like behavior, with an initial drop in $f_g$ at small $\alpha$, followed by a sharp rise. This behavior finds an easy explanation in \reffig{fig:SM_gauge}(a). The fixed-point value $\tilde{G}^{\ast}$\cite{Bonanno:2025tfj} drops sharply for 
$\alpha\lesssim 0.3$ and then stabilizes up to 
$\alpha\approx 2$. As Eqs.~(\ref{eq:fgfinal}), (\ref{eq:fgsharp}) show, 
at seemingly constant $\tilde{G}$, $\alpha$ ends up affecting $f_g$ linearly, hence the sharp rise.

Unlike in the case of minimal matter, with SM content we do not agree quantitatively with the ERG study in Ref.\cite{Riabokon:2025ozw} at $\alpha=0$. The discrepancy of the two results, about a factor of~30, is unlikely to be due entirely to the difference in the adopted regularization procedure, and it may be due, at least in part,  to the different values of the gravitational fixed points $(\tilde{\Lambda}^{\ast},\tilde{G}^{\ast})$ in ERG and PT flow.    

Finally, in \reffig{fig:SM_gauge}(d) we show the $\alpha$-dependence of~$-f_y$. For both choices of the regulator the result is negative and shows a remarkable stability under variations of the gauge fixing parameter at $\alpha\gsim 0.5$. On the other hand, it is also clear that the regulator~$m$ has significant impact on the size of $f_y$. In particular, one can see that $f_y$ is exponentially suppressed at $m\to \infty$ by the larger, negative values of the cosmological constant, cf.~\reffig{fig:SM_gauge}(b). It can make $f_y$ smaller than $f_g$ by one-to-few orders of magnitude.   

%%%%%%%%%%%%%%%%%%%%%%%%%%%%%%%%%%%%%%%%%%%%%%%%%%%%%%%%%%%%%%%%%%%%%%%%%%%%%%%%%%%%%%%%%
\subsection{Asymptotic safety of the matter couplings\label{sec:AS_mat}}

In this subsection, we roughly compare the values obtained for $f_g$ and $f_y$ in the PT renormalization scheme with typical expectations derived from several phenomenological studies of AS in the gauge/Yukawa sector of the SM and beyond. 

It has long been known that a trans-Planckian interactive fixed point for the hypercharge gauge coupling of the SM would provide an elegant and predictive solution to the triviality problem (see, for example, Ref.\cite{Eichhorn:2017lry}). Phenomenological considerations imply that such a solution is obtained if 
the second term on the r.h.s.~of \refeq{eq:betag} cancels the first, \textit{i.e.},  
for a positive $f_g\approx 10^{-2}$ (more precisely, $f_g= 0.0096$ if the SM RG flow is calculated at one loop, in the absence of additional interactions all the way up to the Planck scale, and assuming that gravity decouples sharply at $M_{\textrm{Pl}}=10^{19}\gev$). While significantly smaller $f_g$ values would not be sufficient to tame the UV singularity of the hypercharge gauge coupling, values larger than $10^{-2}$ would enforce asymptotic freedom, rather than safety, with the hypercharge $g_Y\to 0$ 
in the deep trans-Planckian regime.      

Similar considerations apply for the Yukawa couplings of the SM and beyond, although the question of which $f_y$ value can enforce AS in  the Yukawa sector strongly depends on the particular Yukawa coupling and the particular model under consideration.  

%%%%%%%%%%%%%%%%%%%%%%%%%%%%%%%%%%%%%%%%%%%%%%%%%%%%%%%%%%%%%%%%%%%%%%%%%%%%%%%%%%%%%%%%%%%%%%%%%
\setlength\tabcolsep{0.25cm}
\begin{table}[t]
\begin{center}
\begin{tabular}{|c|c|c|c|c|}
\hline
Model & $f_g$ & predictions & $f_y$ & predictions \\
\hline
SM & $f_g=0.0096$ & $g_Y$ safe & $[-10^{-4},10^{-3}]$ & $y_t$ safe \\
\cite{Eichhorn:2017ylw,Eichhorn:2018whv,Kowalska:2022ypk}  & $f_g>0.0096$ & $g_Y$ free & $[-10^{-4},10^{-4}]$ & $y_b$ safe \\
\hline
$B-L$& $f_g=0.0097$ & $g_Y,g_X,g_{\epsilon}$ safe & $ f_y=0.002$ & $y_t,y_{\nu},y_N$ safe\\
 \cite{Kotlarski:2023mmr,Chikkaballi:2023cce}    & $[0.0097,0.05]$ & $g_Y$ free; $g_X,g_{\epsilon}$ safe & $[-0.004,0.005]$ & $y_t$ safe; $y_N$ safe/zero \\
\hline
 SU(6) GUT &  &  &  & $y_t$ safe \\
\cite{Chikkaballi:2025pnw}  &  $f_g\geq 0$ & $g_6$ free & $[0.003,0.04]$ & GUT Yukawa sect.\\
 & & & & safe/zero\\
\hline
\end{tabular}
\caption{Typical values of $f_g$ and $f_y$ required to enforce AS in the SM and particle-physics models of phenomenological interest. $g_Y$ is the hypercharge gauge coupling and $y_t$, $y_b$ are, respectively, the top and bottom quark Yukawa couplings. In the $B-L$ model\cite{Jenkins:1987ue,Buchmuller:1991ce}, $g_X$ and $g_{\epsilon}$ are abelian gauge couplings corresponding, respectively, to the U(1)$_{B-L}$ gauge and the kinetic mixing between U(1)$_Y$ and U(1)$_{B-L}$. $y_{\nu}$ and $y_N$ are, respectively, the Yukawa couplings of the Dirac neutrino to the Higgs boson and of the heavy Majorana neutrino to a U(1)$_{B-L}$-charged scalar SM singlet. In the SU(6) GUT model described in Ref.\cite{Chikkaballi:2025pnw}, $g_6$ is the non-abelian gauge coupling of the GUT group and, in the fifth column, we indicate additional Yukawa couplings connecting the scalar and fermion multiplets of the GUT theory.}
\label{tab:models}
\end{center}
\end{table}
%%%%%%%%%%%%%%%%%%%%%%%%%%%%%%%%%%%%%%%%%%%%%%%%%%%%%%%%%%%%%%%%%%%%%%%%%%%%%%%%%%%%%%%%%%%%%%%%%%%

In \reftable{tab:models} we summarize the values of $f_g$ and $f_y$ that are required to enforce AS 
in the SM and BSM scenarios of phenomenological interest analyzed in the literature. 
For example, it was shown in Ref.\cite{Chikkaballi:2023cce} in the $B-L$ model\cite{Jenkins:1987ue,Buchmuller:1991ce} that, while $f_g>0.0097$ would make the hypercharge gauge coupling asymptotically free, values in the range $0.01\lesssim f_g \lesssim 0.05$ would guarantee that 
the $B-L$ abelian gauge coupling~$g_X$ and kinetic mixing~$g_{\epsilon}$ assume asymptotically safe perturbative values in the trans-Planckian UV. As those fixed points corresponds to irrelevant directions of the RG flow, they provide unique predictions for the corresponding couplings at every scale, even below $M_{\textrm{Pl}}$. In the same model, values in the range $-0.005\lesssim f_y \lesssim 0.005$ guarantee the trans-Planckian safety of the Yukawa coupling~$y_N$, which connects the Majorana neutrinos to a heavy scalar that may trigger electroweak symmetry breaking via a first-order phase transition in the early Universe. Because of this first-order phase transition, the predictions of AS in the $B-L$ model could potentially be tested in future observations of the gravitational-wave background\cite{Chikkaballi:2023cce}. In another example, it was shown in Ref.\cite{Chikkaballi:2025pnw}, in a model based on SU(6) Grand Unified Theory~(GUT), that in presence of trans-Planckian interactive fixed points for the Yukawa couplings of the SU(6) model, a dark matter candidate can unexpectedly emerge after the spontaneous breaking of the GUT gauge group down to the SM, with specific testable properties. For this to occur, $0.003\lesssim f_y\lesssim 0.05$ is required.   

If a coupling admits an interactive fixed points it is dubbed in \reftable{tab:models} as ``safe,'' whereas if it admits a Gaussian fixed point along a relevant direction it is dubbed as ``free.'' If a coupling admits a Gaussian fixed point of the irrelevant type it is dubbed as ``zero.'' We reiterate that zero and safe fixed points that correspond to irrelevant directions of the RG flow provide the predictions of the theory, uniquely determined by the size of the gravitational corrections $f_g$ and $f_y$. We investigate here whether the $f_g$ and $f_y$ computed in the PT scheme can be found, marginally at least, in agreement with the expectations of the models in \reftable{tab:models}. 

%%%%%%%%%%%%%%%%%%%%%%%%%%%%%%%%%%%%%%%%%%%%%%%%%%%%%%%%%%%%%%%%%%%%%%%%%%%%%%%%%%%%%%%%
\begin{figure}[t]
	\centering%
 	    \subfloat[]{%
		\includegraphics[width=0.48\textwidth]
        {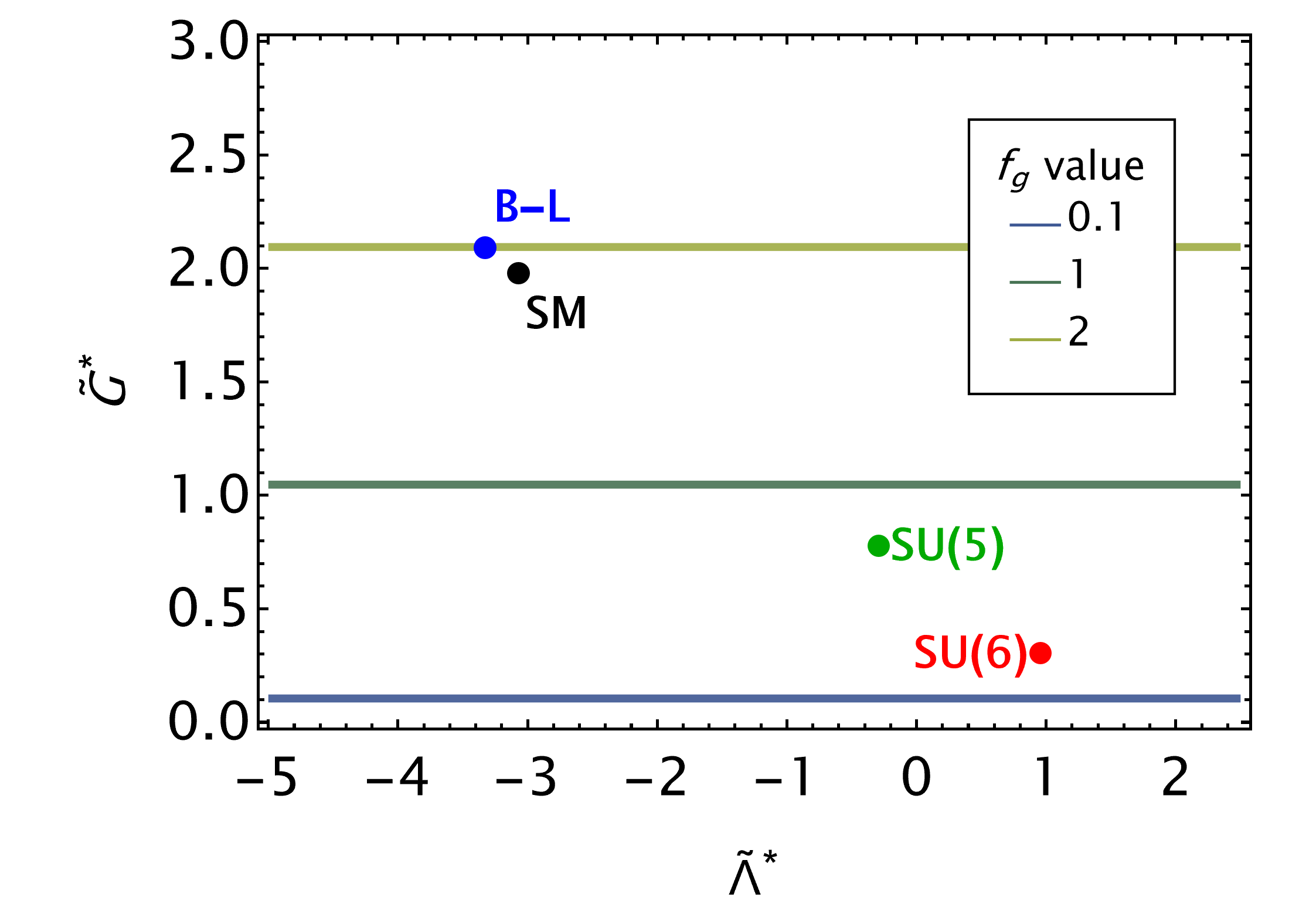}}
  		\hspace{0.2cm}
    	\subfloat[]{%
  		\includegraphics[width=0.48\textwidth]
        {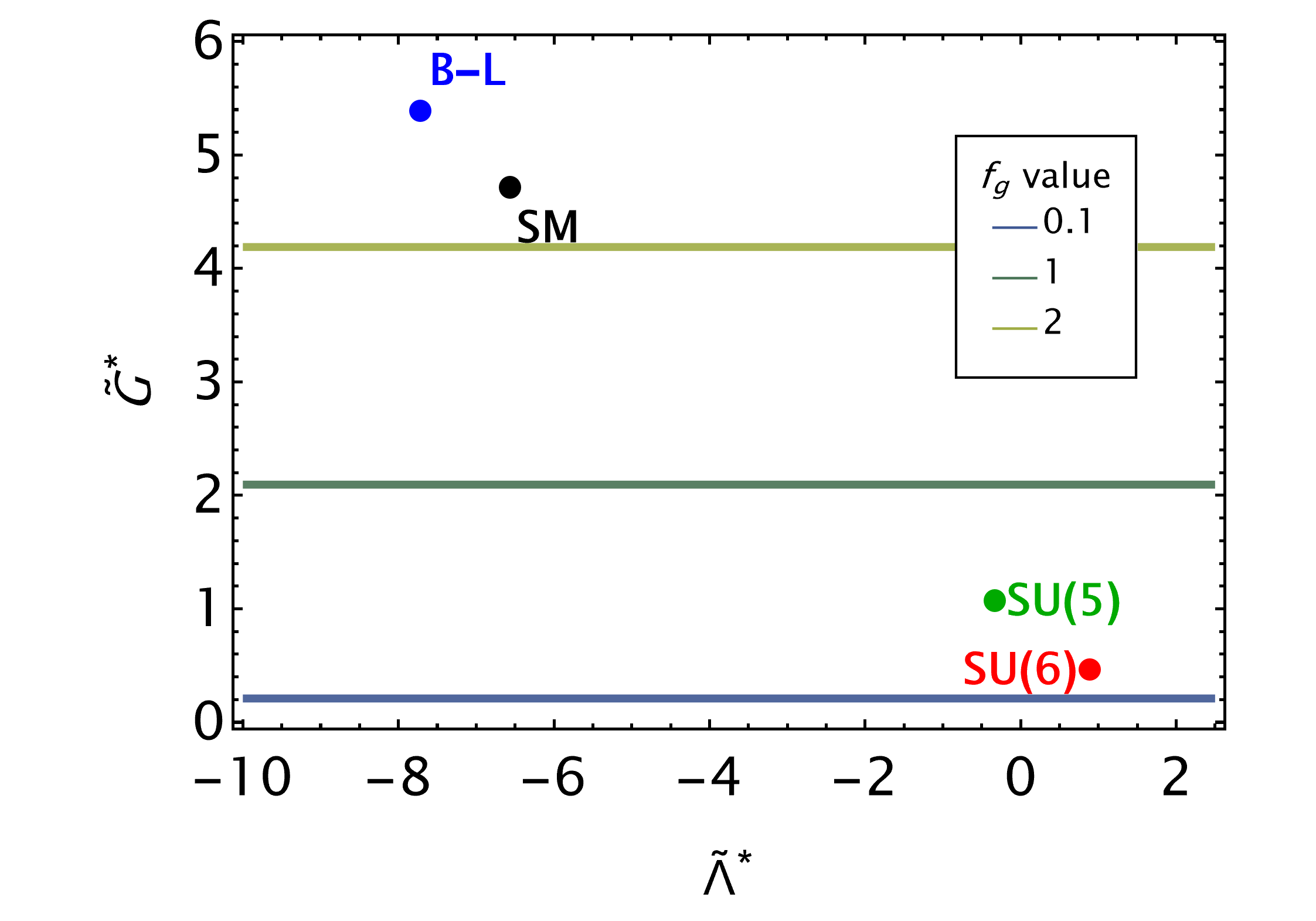}}\\
       	    \subfloat[]{%
		\includegraphics[width=0.48\textwidth]
        {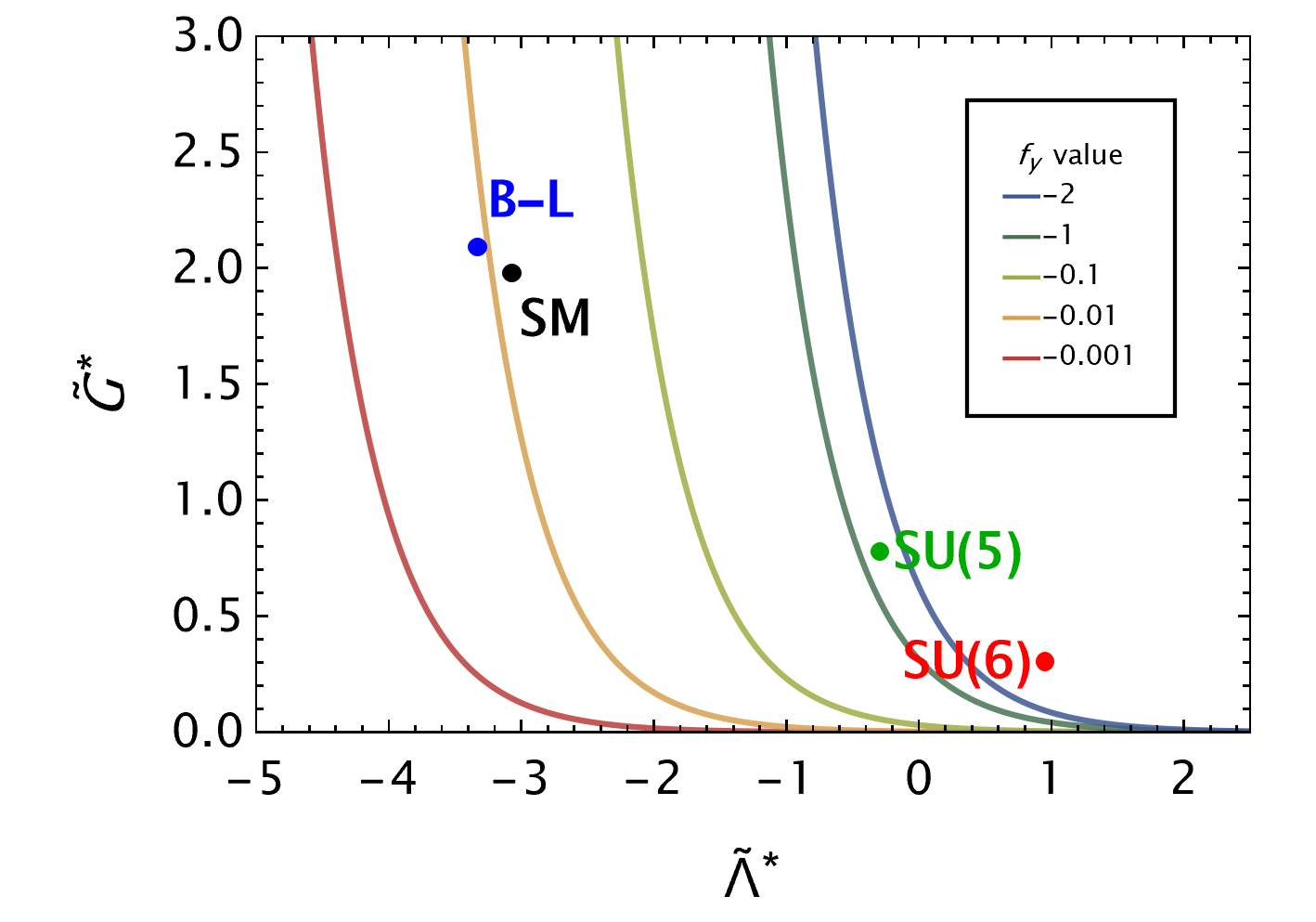}}
  		\hspace{0.2cm}
    	\subfloat[]{%
  		\includegraphics[width=0.48\textwidth]
        {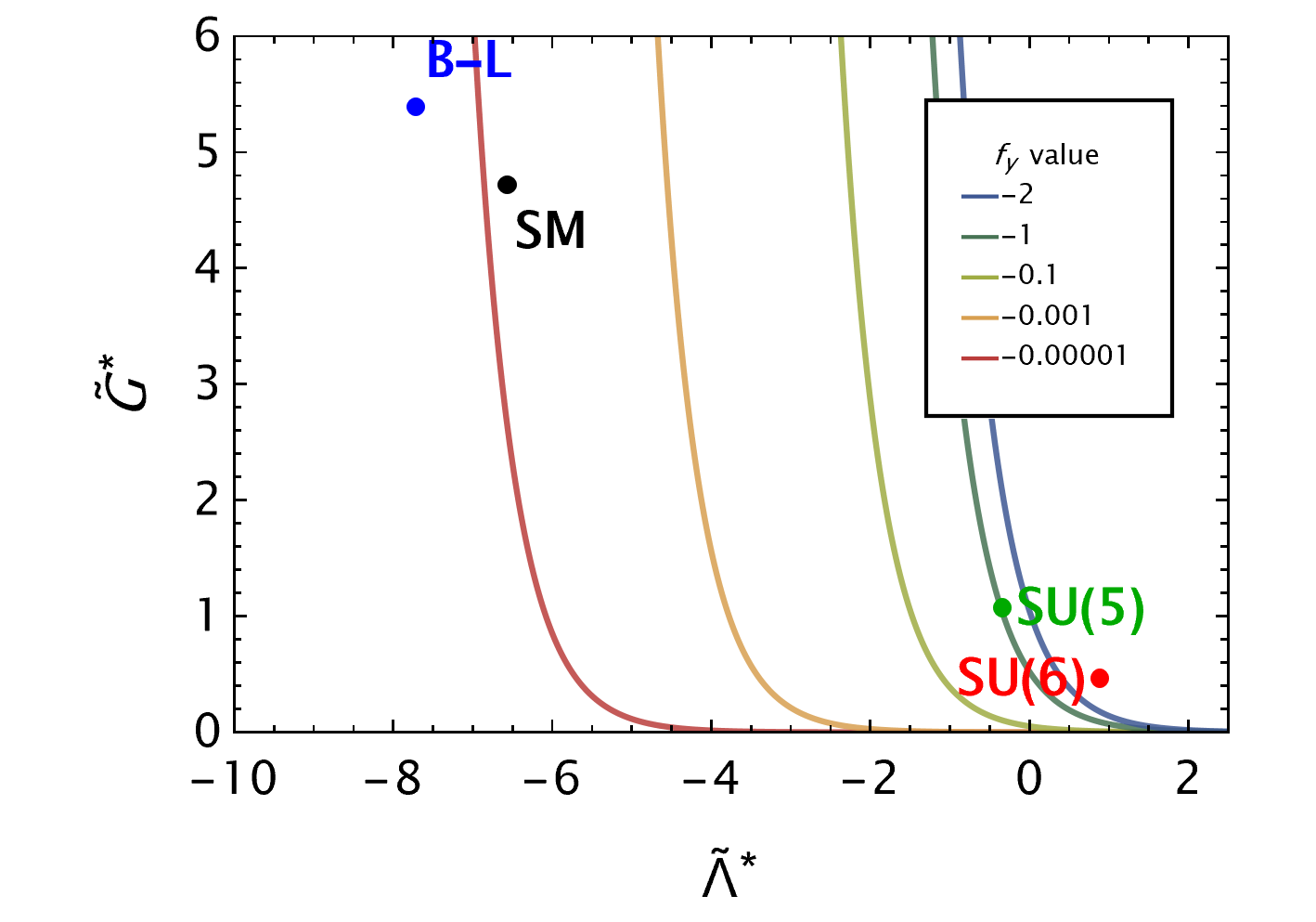}}\\  
\caption{Isocontour lines of $f_g$
in the $(\tilde{\Lambda}^*, \tilde{G}^*)$ plane, computed in the sharp-regulator limit of the proper time flow with gauge-fixing parameter (a)~$\alpha=1$ and (b)~$\alpha=0$. The markers indicate the gravitational UV fixed points for the SM, $B-L$, SU(5), and SU(6) matter content. Isocontour lines of $f_y$ in the $(\tilde{\Lambda}^*, \tilde{G}^*)$ plane, computed in the sharp-regulator limit of the proper time flow with gauge-fixing parameter (c)~$\alpha=1$ and (d)~$\alpha = 0$.}
\label{fig:isocontour}
\end{figure}
%%%%%%%%%%%%%%%%%%%%%%%%%%%%%%%%%%%%%%%%%%%%%%%%%%%%%%%%%%%%%%%%%%%%%%%%%%%%%%%%%%%%%%%%%%%%

In \reffig{fig:isocontour}(a) and \reffig{fig:isocontour}(b) we show, respectively, isocontours of $f_g$ 
in the Feynman~($\alpha=1$) and Landau~($\alpha=0$) gauge,
as a function of the fixed-point values~$\tilde{\Lambda}^{\ast}$, $\tilde{G}^{\ast}$ of the Einstein-Hilbert action. Note that in our derivation $f_g$ does not depend explicitly on $\tilde{\Lambda}$. 
We also show on the plots the fixed-point values corresponding to the models introduced in \reftable{tab:models}, obtained from the zeros of Eqs.~(\ref{eq:betaG}), (\ref{eq:betalam}) in Appendix~\ref{app:grav_beta}. The SM and the $B-L$ model ($N_S=6$, $N_D=24$,  $N_V=13$) feature similar particle content and thus sit close to each other, at the top left corner of the plots. The SU(6) GUT model of Ref.\cite{Chikkaballi:2025pnw}, with $N_S=131$, $N_D=81/2$,  $N_V=35$, sits at the bottom right corner, and we show for completeness also the particle content of vanilla SU(5) GUT, with $N_S=34$, $N_D=24$,  $N_V=24$.  

Qualitatively, the two plots show very similar information, independent of the gauge-fixing choice. 
As was shown in \refsec{sec:SM_mat}, the fixed-point $\tilde{G}^{\ast}$ generates $f_g>1$ in the SM and in its closely related cousin, the $B-L$ model, see Figs.~\ref{fig:SM_gauge}(a) and \ref{fig:SM_gauge}(c). If we take our calculation at face value, a comparison with 
\reftable{tab:models} shows that there is likely no room for a perturbative interactive fixed point with predictivity features in the abelian gauge sector of both models. On the other hand, since the gauge-coupling beta functions admit a Gaussian (zero) fixed point, asymptotic freedom in the abelian gauge sector remains an open possibility. In the latter case, the critical exponent of the corresponding relevant direction would be dominated by the value of $f_g$. In this sense, the observed values of $f_g$ render the PT renormalization scheme entirely consistent with the expectations of the asymptotically safe SM\cite{Pastor-Gutierrez:2022nki,Riabokon:2025ozw,deBrito:2025nog}. 

In GUT based models SU(5) and SU(6) we observe, due to their richer particle content, smaller values of $\tilde{\Lambda}^{\ast}$, $\tilde{G}^{\ast}$ with respect to the SM. While reducing $\tilde{G}^{\ast}$ can bring $f_g$ down by up to an order of magnitude with respect to the SM (or the $B-L$ model), the gravitational contribution to the gauge sector remains substantial in GUTs. This is not really a problem, as the unified gauge coupling is already, in most cases, asymptotically free in a GUT.

The conclusions drawn above need a word of caution. They are derived under the approximations used in this work, based on the fixed points of the standalone Einstein-Hilbert action\cite{Bonanno:2025tfj}. As we have discussed in \refsec{sec:gauge}, an extended truncation of the action or the implementation of RG improvement may 
produce a reduced value of $f_g$ for these models, for example by reintroducing an explicit dependence on the cosmological constant. 

One can follow similar lines for the discussion of the Yukawa sector and in this regard we present in  \reffig{fig:isocontour}(c) and \reffig{fig:isocontour}(d), respectively, isocontours of $f_y$ 
in the Feynman~($\alpha=1$) and Landau~($\alpha=0$) gauge,
as a function of the fixed-point values~$\tilde{\Lambda}^{\ast}$, $\tilde{G}^{\ast}$ of the Einstein-Hilbert action. 
It is well known that the negative values typically obtained for the leading gravity correction to the Yukawa coupling in the ERG (and, now, PT scheme) are phenomenologically problematic at the Gaussian fixed point of the asymptotically safe SM. In fact, at the Gaussian fixed point the negative sign will dominate the critical exponent and make the Yukawa coupling irrelevant (we referred to solutions of this class as ``zero'' in \reftable{tab:models}, as they are protected from appearing at any sub-Planckian scale). Additionally, \reftable{tab:models} also shows that negative values of $f_y$ 
are incompatible with the viability of the dark matter predictions associated with AS in the Yukawa sector of the SU(6) model introduced in Ref.\cite{Chikkaballi:2025pnw} -- which require $f_y>0$ -- and with phenomenologically viable Yukawa couplings in generic GUT models. 

In general, when taking the $f_y$ calculation at face value, we observe that a large, negative $\tilde{\Lambda}^{\ast}<0$ can induce a significant suppression of the gravitational contribution to the running Yukawa coupling. This leaves in principle the door open, in the SM and the $B-L$ model, to the existence of perturbative, predictive fixed points of the Yukawa couplings that might lead to observable features. For example, 
$f_y\approx -10^{-4}$ could imply that either the top or the bottom Yukawa coupling of the SM (or both\cite{Eichhorn:2025sux})
are a prediction of the scale-invariant UV completion. In the $B-L$ model, $f_y\approx -10^{-3}$ could imply that the size of the Mayorana-neutrino Yukawa coupling becomes a prediction, with the enticing possibility of discriminating different scenarios through gravitational-wave signals\cite{Chikkaballi:2023cce}.
That said, one should remain aware that the predictions drawn here are intended to be qualitative indications that need to withstand further quantitative scrutiny. This is because, on the one hand, when the tadpole contribution to the Yukawa coupling is highly suppressed by a non-perturbative $\tilde{\Lambda}^{\ast}<0$, the impact of extended truncations of the gravitational action and of additional diagrams in the Yukawa-coupling computation may become substantial. But, perhaps more importantly, in light of the fact, observed in a recent state-of-the art analysis of the Yukawa sector in the ERG\cite{deBrito:2025nog}, that a next-to-leading order resummation of operators contributing to the Yukawa coupling can flip the sign of the critical exponent at the Gaussian fixed point. This
happens even in the parameter-space region with $|\tilde{\Lambda}^{\ast}|\ll 1$, where the negative sign and approximate size of the gravitational contribution have been established at the leading order under different choices of regulator and gauge fixing. Given the qualitative similarities between PT flow and ERG we observe in our study, it is reasonable to expect that the higher-order sign flip may be as well a feature of the former, as it is of the latter. We leave a more detailed analysis of this issue for future work.

%%%%%%%%%%%%%%%%%%%%%%%%%%%%%%%%%%%%%%%%%%%%%%%%%%%%%%%%%%%%%%%%%%%%%%%%%%%
\section{Summary and conclusions}\label{sec:con}
%%%%%%%%%%%%%%%%%%%%%%%%%%%%%%%%%%%%%%%%%%%%%%%%%%%%%%%%%%%%%%%%%%%%%%%%%%%%

In this work, we have employed the Schwinger PT flow equation to derive the quantum gravity contributions to the beta functions of the gauge and 
Yukawa coupling of a matter theory. Following common practice in the literature on AS, we have dubbed these gauge group- and flavor-independent corrections to the running couplings as $f_g$ and $f_y$. Working with a flat background metric and in the Einstein-Hilbert truncation, our results show explicitly the gauge-fixing and PT-regulator dependence of $f_g$ and $f_y$. To quantify the sensitivity of our results to these unphysical parameters, and facilitate a direct comparison with recent determinations of $f_g$ and $f_y$ in the ERG, our results are computed at the interactive fixed point of the quantum gravity action, which was derived by one of us in the PT scheme in a recent study.     

We found remarkable agreement with existing, leading-order ERG determinations, especially in the case of a minimal matter content (one vector field, one Dirac fermion, one real scalar) where, given equivalent choices for the truncation and background metric, we obtain comparable values of $f_g$ and $f_y$ at the Reuter-like UV fixed point of the Newton and cosmological constants. These results make a strong case that -- despite AS being typically associated with the functional RG -- the same asymptotically safe behavior of the gravity-matter system emerges independently of the specific choice of regularization technique. 

On a less positive note, our PT flow calculation confirms the leading negative sign of the gravitational tadpole contribution to the Yukawa coupling, which was long observed in the ERG and persists for $\Lambda$ resummed to all orders in the heat kernel expansion. Such negative sign is phenomenologically problematic at the Gaussian fixed point of the asymptotically safe SM, where it dominates the critical exponent and consequently renders the Yukawa coupling irrelevant. Given our findings, it appears that a next-to-leading order analysis, featuring the full resummation of off-diagonal operators along the lines of Ref.\cite{deBrito:2025nog}, becomes necessary to prove the phenomenological viability of the Yukawa sector in PT flow as well.

The above considerations reflect in our phenomenological analysis, where we have compared the determinations of $f_g$ and $f_y$ with typical expectations for their sign and size
extracted in heuristic studies of the observational properties of the SM and/or BSM scenarios embedded in an asymptotically safe, quantum scale invariant trans-Planckian UV completion. While the size of $f_g$ at the gravity fixed point is entirely consistent with the ansatz of the asymptotically safe SM -- where, by construction, the gauge couplings are all asymptotically free and stem from a relevant Gaussian UV fixed point\cite{Pastor-Gutierrez:2022nki} -- accommodating interactive fixed points of the irrelevant kind for the matter sector -- which would spawn some of the spectacular signatures predicted in the literature in presence of a properly scale-invariant UV completion -- appears to be a real challenge in this scheme, at least under the assumptions adopted in this work.

As for the Yukawa coupling, the leading negative sign of $f_y$ prevents the appearance of relevant Gaussian fixed points in the asymptotically safe SM but some interesting predictions may instead stem from irrelevant interactive fixed points, which appear in new physics scenarios like, \textit{e.g.}, the $B-L$ model. Needless to say,
such indications are still very qualitative and need to be anchored in firmer quantitative ground.  
To this regard, our study can be extended along different directions. It would be interesting to compute the PT flow of $f_g$ and $f_y$ in concomitance with the flow of the gravitational action, so to estimate the respective back reaction. Extended truncations of the gravity and matter actions, non-flat backgrounds, higher order effects via RG improvement, and the aforementioned next-to-leading order resummation of operators, are very likely to have an impact on the sign and size of the trans-Planckian matter beta functions and may generate values more in line with expectations of predictivity arising in the SM and BSM scenarios. Overall, we think our study reinforces the generality of the AS ansatz and provides a possible first step towards a systematic adaptation of the results obtained in the context of the ERG to alternative regularization schemes.

%%%%%%%%%%%%%%%%%%%%%%%%%%%%%%%%%%%%%%%%%%%%%%%%%%%%%%%%%%%%%%%%%%%%%%%%%%%%%%%%
\newpage
\begin{center}
\textbf{ACKNOWLEDGMENTS}
\end{center}
We would like to thank M.~Reichert for asking a question that led us to identify an error in our previous calculation of $f_y$.
GG wishes to thank the National Centre for Nuclear 
Research in Warsaw for the kind hospitality during the 
initial stages of this project. GG and DZ are grateful 
to G.~Oglialoro and E.~M.~Glaviano for enlightening 
discussions. KK and EMS would like to thank M.~Schiffer 
for stimulating discussions. DR is supported by the 
Estonian Research Council grants TARISTU24-TK10, 
TARISTU24-TK3, and the CoE grant TK202 “Foundations of 
the Universe.” EMS is supported in part by the National 
Science Centre (Poland) under the research Grant 
No.~2020/38/E/ST2/00126.  
\bigskip
%%%%%%%%%%%%%%%%%%%%%%%%%%%%%%%%%%%%%%%%%%%%%%%%%%%%%%
\newpage
%%%%%%%%%%%%%%%%%%%%%%%%%%%%%%%%%%%%%%%%%%%%%%%%%%%%%%%%%%%%%%%%%%%%%%%%%%%%%%%%%%%%%%%%%%%%%%%%%%%%%%
\begin{figure}[t]
	\centering%
    	\subfloat[]{%
  		\includegraphics[width=0.48\textwidth]
        {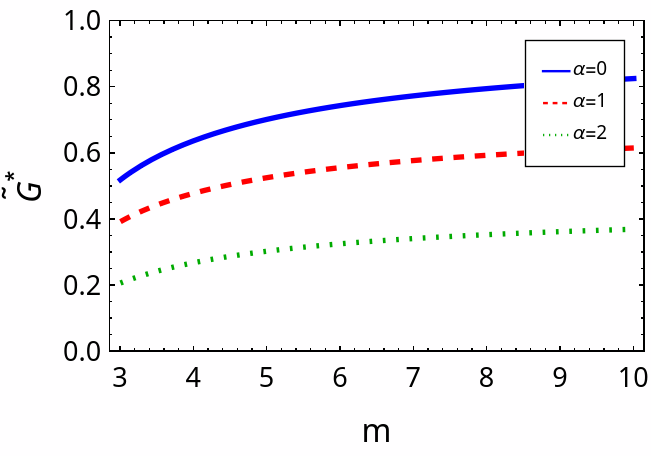}}
                  \hspace{0.2cm}
    	\subfloat[]{%
  		\includegraphics[width=0.48\textwidth]
        {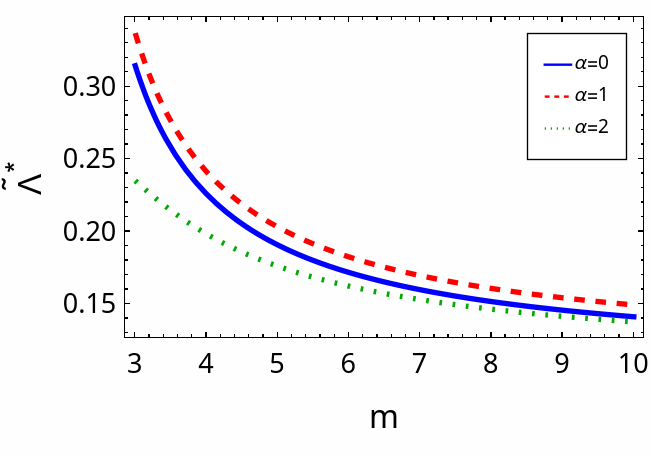}}\\
        \subfloat[]{%
  		\includegraphics[width=0.48\textwidth]
        {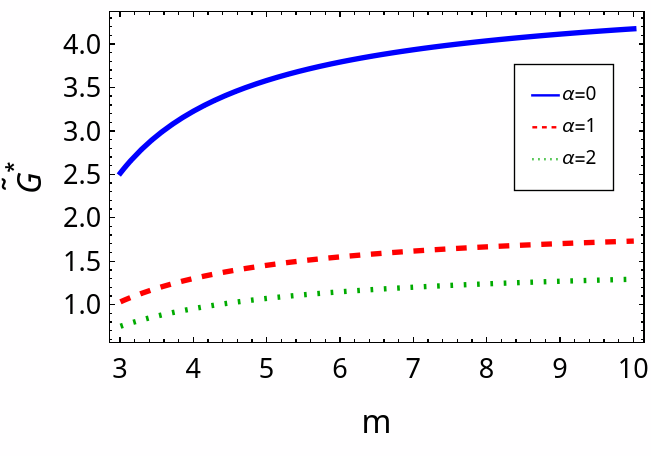}}
                  \hspace{0.2cm}
    	\subfloat[]{%
  		\includegraphics[width=0.48\textwidth]
        {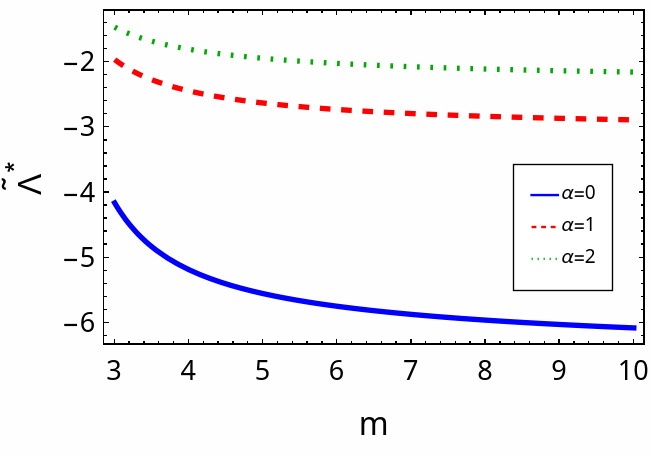}}\\    
\caption{Regulator dependence of the fixed-point values of the Einstein-Hilbert action, (a)~$\tilde{G}^\ast$ and (b)~$\tilde{\Lambda}^\ast$ in the minimal matter model for three values of the gauge-fixing parameter: $\alpha=0$~(solid blue),  $\alpha=1$~(dashed red), and $\alpha=2$~(dotted green). (c) Same as~(a), in the SM. (d) Same as~(b), in the SM.}
\label{fig:gravitym}
\end{figure}
%%%%%%%%%%%%%%%%%%%%%%%%%%%%%%%%%%%%%%%%%%%%%%%%%%%%%%%%%%%%%%%%%%%%%%%%%%%%%%%%%%%%%%%%%%%%%%%%%%%%%%

\begin{appendix}

\section{Fixed points of the gravity sector\label{app:grav_beta}}

In this appendix, we recall the beta functions of the dimensionless parameters~$\tilde{\Lambda}$ and $\tilde{G}$ of the Einstein-Hilbert action, which were computed in Ref.\cite{Bonanno:2025tfj}.
The matter content is added to the beta functions via the number of real scalar 
($N_S$), Dirac fermion ($N_D$), and vector ($N_V$) fields. One obtains
\begin{equation}\label{eq:betaG}
    \beta_{\tilde{G}} = 2\, \tilde{G} + \eta_{G} \tilde{G}+\frac{\tilde{G}^2}{6\pi}
    \left(2N_D + N_S - 4N_V\right),
\end{equation}
\begin{equation}\label{eq:betalam}
    \beta_{\tilde{\Lambda}} = \left(\eta_G-2\right)\tilde{\Lambda}+\Phi + \frac{\tilde{G}}{4\pi}\left(N_S - 4N_D + 2N_V\right) + \frac{\tilde{G}\tilde{\Lambda}}{6\pi}\left(2N_D + N_S - 4N_V\right)\,,
\end{equation}
where\cite{Bonanno:2025tfj} 
\begin{multline}\label{eq:etagrav}
    \eta_G=\frac{\tilde{G}}{12 \pi  (m-1)} \Bigg[\frac{2 m^m \left(\frac{4 \tilde{\Lambda} }{\alpha -3}+m\right)^{-m} \left(4 \tilde{\Lambda} +\left(\alpha -3\right) m\right)}{\alpha -3}\\+\frac{2 \left(7 \alpha -9\right) m^m \left(\frac{4 \alpha \tilde{\Lambda} }{\alpha -3}+m\right)^{-m} \left(4 \alpha \tilde{\Lambda} +\left(\alpha -3\right) m\right)}{\left(\alpha -3\right)^2}\\-6 \left(3\alpha -2\right) m^m \left(m-2 \alpha  \lambda \right)^{1-m}-50 m^m \left(m-2 \lambda \right)^{1-m}+\frac{4 \left(27-7 \alpha \right) m}{\alpha -3}\Bigg],
\end{multline}
and 
\begin{multline}
    \Phi=-\frac{\tilde{G}\, \Gamma (m-2) }{4 \pi  \Gamma (m)}\Bigg[-2 m^m \left(\frac{4 \tilde{\Lambda} }{\alpha -3}+m\right)^{2-m}-6 m^m \left(m-2 \alpha \tilde{\Lambda} \right)^{2-m}\\-10 m^m \left(m-2 \tilde{\Lambda}
 \right)^{2-m}-2 m^m \left(\frac{4 \alpha \tilde{\Lambda} }{\alpha -3}+m\right)^{2-m}+16 m^2\Bigg].
\end{multline}

In \reffig{fig:gravitym}(a) we show the dependence of the fixed point value $\tilde{G}^{\ast}$ on the regulator parameter~$m$, assuming minimal matter content ($N_S=N_D=N_V=1$),
for three different values of the gauge fixing parameter~$\alpha$. The dependence of $\tilde{\Lambda}^{\ast}$ on the regulator parameter~$m$ for the same three values of~$\alpha$ is shown in 
\reffig{fig:gravitym}(b). 

Figures~\ref{fig:gravitym}(c) and \ref{fig:gravitym}(d) show the $m$-dependence of $\tilde{G}^{\ast}$
and $\tilde{\Lambda}^{\ast}$, respectively, 
assuming the SM matter content ($N_S=4$, $N_D=45/2$, $N_V=12$).  

\end{appendix}
%%%%%%%%%%%%%%%%%%%%%%%%%%%%%%%%%%%%%%%%%%%%%%%%%%%%%%%
\bibliographystyle{utphysmcite}
\bibliography{mybib}

\end{document}